\documentclass[a4paper,11pt]{article}
\usepackage{graphicx,amssymb,amstext,amsmath,mathabx}
\usepackage{color, yfonts, tcolorbox}
\usepackage{floatflt}
\usepackage{tikz,pgfplots}
\pgfplotsset{compat=1.3}
\usetikzlibrary{spy}
\usepackage{hyperref}
\hypersetup{
    colorlinks=true,
    linkcolor= blue,
    filecolor=magenta,      
    urlcolor=blue,
}

\definecolor{cbl}{rgb}{0,0,1}                % bleu\

\usepackage{mathtools}

\newcommand{\bc}{\begin{center}}
\newcommand{\ec}{\end{center}}
\def\ba#1{\begin{array}{#1}\displaystyle}
\newcommand{\ea}{\end{array}}

\newcommand{\beq}{\begin{equation}}
\newcommand{\eeq}{\end{equation}}
\newcommand{\beqa}{\begin{eqnarray}}
\newcommand{\eeqa}{\end{eqnarray}}

\newcommand{\bi}{\begin{itemize}}
\newcommand{\ei}{\end{itemize}}
\newcommand{\p}{\partial}

\newcommand{\ep}{\epsilon}
\newcommand{\varep}{\varepsilon}

\newcommand{\Tr}{{\rm Tr}}

\begin{document}
\begin{titlepage}
\vspace{0.2cm}
\begin{center}

{\large{\bf{On the Hydrodynamics of Unstable Excitations}}}

\vspace{0.8cm} 
{\large \text{Olalla A. Castro-Alvaredo${}^{\heartsuit}$, Cecilia De Fazio${}^{\diamondsuit}$, Benjamin Doyon${}^{\spadesuit}$, and Francesco Ravanini${}^{\clubsuit \, {\LARGE \bullet}}$}}

\vspace{0.8cm}
{\small
{\small ${}^{\heartsuit\,\diamondsuit}$} Department of Mathematics, City, University of London, 10 Northampton Square EC1V 0HB, UK\\
\vspace{0.2cm}
${}^{\spadesuit}$  Department of Mathematics, King's College London, Strand WC2R 2LS, UK \\
\vspace{0.2cm}
${}^{\clubsuit}$  Dipartimento di Fisica, Universit\`aÂ  di Bologna, Via Irnerio 46, I-40126 Bologna, Italy \\
\vspace{0.2cm}
${}^{{\LARGE \bullet}}$  INFN, Sezione di Bologna, Via Irnerio 46, I-40126 Bologna, Italy
}
\end{center}
\medskip
\medskip
\medskip
\medskip
The generalized hydrodynamic (GHD) approach has been extremely successful in describing the out-of-equilibrium properties of a great variety of integrable many-body quantum systems. It naturally extracts the large-scale dynamical degrees of freedom of the system, and is thus a particularly good probe for emergent phenomena. One such phenomenon is the presence of unstable particles, traditionally seen via special analytic structures of the scattering matrix. Because of their finite lifetime and energy threshold, these are especially hard to study. In this paper we apply the GHD approach to a model possessing both unstable excitations and quantum integrability. The largest family of relativistic integrable quantum field theories known to have these features are the homogeneous sine-Gordon models.
%These factorized-scattering models possess unstable particles and break parity invariance.
We consider the simplest non-trivial example of such theories and investigate the effect of an unstable excitation on various physical quantities, both at equilibrium and in the non-equilibrium state arising from the partitioning protocol. The hydrodynamic approach sheds new light onto the physics of the unstable particle, going much beyond its definition via the analytic structure of the scattering matrix, and clarifies its effects both on the equilibrium and out-of-equilibrium properties of the theory. Crucially, within this dynamical perspective, we identify unstable particles as finitely-lived bound states of co-propagating stable particles of different types, and observe how stable populations of unstable particles emerge in large-temperature thermal baths.

\vspace{1cm}

\medskip

\vfill

\noindent 
${}^{\heartsuit}$ o.castro-alvaredo@city.ac.uk\\
${}^{\diamondsuit}$ cecilia.de-fazio.2@city.ac.uk\\
${}^{\spadesuit}$ benjamin.doyon@kcl.ac.uk\\
${}^{\clubsuit}$ francesco.ravanini@bo.infn.it\\

\hfill \today

\end{titlepage}
\section{Introduction}
Over the past decade, the out-of-equilibrium dynamical properties of many-body quantum systems have been extensively studied \cite{Eisert}. The interface between this rather general problem and integrable systems has been particularly rich in new results. Since the famous Quantum Newton's Cradle experiment \cite{kinoshita} it has been known that the role of integrability (i.e.~the presence of a large number of conservation laws) in one-dimensional systems has dramatic implications for the dynamics of such models. In particular, it is now well-understood that the dynamics of one-dimensional integrable models following a quantum quench  is described by a Generalized Gibbs Ensemble (GGE) \cite{Rigol}, that is, a partition function involving all local and quasi-local conserved charges in the system
\beq
{\mathcal Z}= \Tr\left(e^{-\sum_{i} \beta_i Q_i}\right)\,.
\label{gge}
\eeq
Therefore integrable systems do not {\it{thermalize}} in the usual sense but they do {\it{relax}} towards a GGE.  In particular, the role of quasi-local and semi-local conserved quantities in the GGE has been the subject of a lot of investigation \cite{failure,Prosen1, Prosen2, Prosen3, Prosen4, doyon2017}, paving the way to a comprehensive understanding for the most paradigmatic integrable spin chain model, the spin-$\frac{1}{2}$ XXZ chain \cite{ilinardo}.  A good summary of the main results up to 2016 is provided in the special issue \cite{CEM}. 

The same year of 2016 saw the solution of a related problem. This was the full understanding of how to compute dynamical quantities in non-equilibrium steady states and non-stationary settings, by employing a (generalized) hydrodynamic approach (GHD) \cite{ourhydro,theirhydro}. The basic idea is that hydrodynamics {\it emerges} as a consequence of local entropy maximization on individual fluid cells containing sufficiently large numbers of quasi-particles. Technically, this is the assumption that averages of local quantities tend uniformly enough, at large times, to averages evaluated in GGEs with space-time dependent potentials ${\beta}_i(x,t)$ in (\ref{gge}). Physically, this is a consequence of separation of scales. It is worth noting that the development of GHD was made possible in particular by the evaluation of exact expectation values of currents in GGEs, derived in \cite{ourhydro} within QFT and numerically checked in \cite{theirhydro} in quantum chains; this particular aspect has received a lot of attention afterwards, with increasingly rigorous and general derivations \cite{cur1,cur2,cur3,dNBD2,cur4,cur5,cur6,cur7,cur8,cur9}.

Let us now suppose that we engineer an out-of-equilibrium set up by employing the partitioning protocol (see e.g.~Fig.~1 in \cite{ourhydro}). This means that we consider two separate systems, each characterized by a particular steady state and set of generalized inverse temperatures $\underline{\beta}=\{\beta_i\}$. In the partitioning protocol the two systems are put into contact at time $t=0$. The presence of multiple conserved quantities gives rise to ballistic transport, meaning that, after a transient period, steady state currents flowing between the right and left sub-systems emerge; see the reviews \cite{BDReview,VMReview}. GHD provides a method  to compute such currents by combining the hydrodynamic principle, generalized to infinitely many conservation laws, with an effective description of quasi-particles readily available for integrable models. For quantum field theories (QFTs) such a description is known as the thermodynamic Bethe ansatz (TBA) \cite{tba1,tba2} and it was generalized to GGEs in \cite{Mossel}. The resulting mathematical procedure is based on the solution of a set of coupled nonlinear integral equations, usually carried out numerically, whose sole inputs are: the one-particle eigenvalues of all conserved charges involved in the GGEs characterizing the original left and right systems, the two-particle scattering matrix of the QFT, and the (stable) particle spectrum of the original theory.  Since the original proposals \cite{ourhydro,theirhydro} a plethora of generalizations have been developed, such as the inclusion of force terms \cite{DY,bastianello2019Integrability,BasGeneralised2019}, diffusive and higher corrections \cite{dNBD,GHKV2018,dNBD2,FagottiLocally}, noise \cite{bastianello2020noise}, integrability breaking terms \cite{CaoTherm18,vas19,DurninTherma2020}, and much more. There is now even experimental evidence that GHD provides a better description of transport in an atom chip than conventional hydrodynamics \cite{chippy}. A pedagogical overview is provided in the lecture notes \cite{benreview}.

A situation that has hitherto escaped attention is the inclusion of unstable excitations in the theory under consideration. In this paper we partly fill this gap by considering a very simple example where the effect of unstable excitations on particular steady-state currents and densities, and on the effective velocities of stable modes, can be well understood. We focus on a very simple integrable QFT known as the $SU(3)_2$-homogeneous sine-Gordon (HSG) model \cite{hsg, ntft, FernandezPousa:1997iu, smatrix}. This is a theory whose spectrum contains two stable particles of the same mass. The two-particle scattering matrix has a pole in the unphysical sheet of non-vanishing real and imaginary parts. This can be interpreted as the creation of an unstable particle, with a finite decay width and mass that can be computed from the usual Breit-Wigner formula.  The theory also has the additional interesting feature of breaking parity invariance.

The presence of an unstable particle and the absence of parity invariance have interesting consequences, which are brought to light most clearly using the dynamical description offered by GHD.  For instance, due to parity breaking, even if total currents are vanishing at equilibrium, the individual contribution to the currents (be it of energy or particles) of each particle type does not vanish. As we explain via the ``flea-gas" picture behind GHD \cite{DYC}, there are natural right-movers and left-movers. Most interestingly, the nontrivial patterns of particle densities and effective velocities developing as the temperature changes allow us to obtain a clear picture behind the formation of unstable particles at energies beyond their threshold. We observe that fundamental particles separate into various groups: co-moving pairs of particles of opposite types, interpreted as finitely-lived bound states and identified with the unstable particle of the spectral theory; and separate freely propagating fundamental particles, identified as residual free fermions. Crucially, at high temperatures, the population of unstable particles reaches a stable proportion: while the particles decay, their population is continuously replenished thanks to the high energy of the state. This picture explains the structure of all quantities evaluated in the out-of-equilibrium state. These groups of particles contribute as separate degrees of freedom to the theory, and this gives a clear interpretation of the total central charge of the large-energy, UV fixed point. In this way, we show how GHD both sheds new light on equilibrium properties of unstable particles, and explains their out-of-equilibrium behaviours.

\medskip

This paper is organized as follows:  In Section 2 we introduce the model, the thermodynamic Bethe ansatz approach and the main principles of the GHD approach in the context of integrable quantum field  theory. In Section 3 we present and discuss numerical results for energy current and density, particle current and density, and effective velocities at equilibrium. In section 4 we present results for the same quantities out-of-equilibrium, starting from two thermal baths.  In section 5 we discuss the application of our results to the problem of determining the energy current and density generated when connecting two conformal field theories of different central charges. We conclude in Section 6.
Appendix A discusses results for the position of the discontinuity of the occupation numbers in the out-of-equilibrium steady state. In Appendix B we review the main features of our numerical algorithm.

\section{Introducing the Model and Main Techniques}
\subsection{The Model}
The family of HSG-models provides one of the few examples where integrability and 
the presence of unstable excitations are successfully combined. These models were first studied in a series of papers in the late 90s where their classical and quantum integrability were established \cite{hsg,ntft}, the particle spectrum studied  \cite{FernandezPousa:1997iu}, and a diagonal scattering matrix proposed \cite{smatrix}. The scattering matrix was then tested extensively by employing the TBA \cite{tba1,tba2}  and the form factor approach \cite{KW,SmirnovBook}. In particular, in this work we will only consider the simplest example of this family of theories, known as the $SU(3)_2$-homogeneous sine-Gordon model. The model may be seen as a massive perturbation of a critical Wess-Zumino-Novikov-Witten model \cite{WZNW1,WZNW2,WZNW3,WZNW4,WZNW5} associated to the coset $SU(3)_2/U(1)^2$, where the subindex 2 in the numerator is a parameter of the model called the level. In general, it is possible to define HSG-models associated to all cosets $G_k/U(1)^{r_g}$ where $G$ is some simply-laced algebra, $k$ is the level (an integer), and $r_g$ is the rank of $G$.
The TBA of this and other models in the same family was studied in detail in \cite{ourtba,CastroAlvaredo:2002nv,Dorey:2004qc} and the form factors of local operators constructed in \cite{CastroAlvaredo:2000em,CastroAlvaredo:2000nk}. The effect of the presence of unstable particles in the RG-flow of several quantities was also explored using form factor techniques in \cite{CastroAlvaredo:2000ag,CastroAlvaredo:2000nr}. 

The $SU(3)_2$-homogeneous sine-Gordon model is an integrable QFT with a two-particle spectrum. We will denote the particles by $\pm$. The scattering matrices are diagonal and simply given by:
\beq\label{Smatrix}
S_{\pm\pm}(\theta)=-1, \qquad S_{\pm\mp}(\theta)=\pm \tanh\frac{1}{2}\left(\theta\pm \sigma-\frac{i\pi}{2}\right)\,,
\eeq
where $\sigma$ is a free parameter of the theory. An interesting feature of this theory (and others in the same family) is parity breaking, namely $S_{+-}(\theta)\neq S_{-+}(-\theta)$.  In addition we have that 
\beq
\lim_{|\sigma| \rightarrow \infty} S_{\pm\mp}(\theta)=1\,,
\label{free}
\eeq
 which means that in this limit parity symmetry is restored and the theory may be seen as two independent, mutually commuting free Majorana fermions. An important consequence of this property is that the behaviour of any quantity we compute at or out of equilibrium is always identical to that of a pair of free  fermions, as long as $|\sigma|$ is large enough and we consider temperatures that are sufficiently small relative to the value of $|\sigma|$, so as not to excite states of large rapidities. This feature constitutes a very useful benchmark and consistency check for our numerics. 
 
For finite $\sigma$, the theory is interacting and the scattering amplitudes $S_{\pm\mp}(\theta)$ have a pole outside the physical sheet at $\theta=\mp\sigma-\frac{i\pi}{2}$, in the strip $-\pi \leq \mathrm{Im}(\theta)\leq 0$. As discussed in \cite{CastroAlvaredo:2000ag} from the Breit-Wigner formula it follows that, for this particular $S$-matrix, assuming the two particles $\pm$ have the same mass $m$, 
\beq
{M}^2=m^2(1+\cosh\sigma)\, \qquad \mathrm{and}\qquad \Gamma^2=4 m^2 (-1+\cosh\sigma)\,,
\eeq
where ${M}$ is the mass of the unstable particle and $\Gamma$ its decay width. Therefore
\beq
{M}\sim \frac{1}{\sqrt{2}} m e^{\frac{|\sigma|}{2}} \quad \mathrm{and} \quad  \Gamma\sim \sqrt{2} m e^{\frac{|\sigma|}{2}}\, \quad \mathrm{for} \quad |\sigma|\gg 0\,.
\eeq
Thus, the larger $|\sigma|$ is, the more massive and short-lived the unstable excitation becomes.

Intuitively, the excitation is considered an unstable bound state of the otherwise two free Majorana fermion species. For $|\sigma|$ large, one would then expect a clear separation of energy scales. That is, as mentioned, at small temperatures compared to the scale set by the mass $M$ and decay width $\Gamma$, the physics is dominated by the two free fermions as unstable bound states decay quickly. In contrast, at large temperatures with respect to this scale, there is enough energy for a finite proportion of particles to be found within bound states, which re-populate fast enough. At large temperatures, the unstable particle has nontrivial, large-scale effects. It is one goal of this paper to obtain a clearer, dynamical picture of these effects, and to identify the unstable particle in a more physically clear fashion than the use of the Breit-Wigner formula.

In our work we will use a logarithmic scale for temperatures (i.e.~we typically plot against $\log \frac{\beta}{2}$ where $\beta$ is some inverse temperature). Thus $\log \frac{\beta}{2}\approx -\frac{|\sigma|}{2}$ is the value which signals the onset of the unstable particle.
Without loss of generality, we choose
\beq
 \sigma>0\,,
 \eeq
for the remainder of this paper. 

{\begin{floatingfigure}[h]{7.5cm} 
 \begin{center} 
\includegraphics[width=7.8cm]{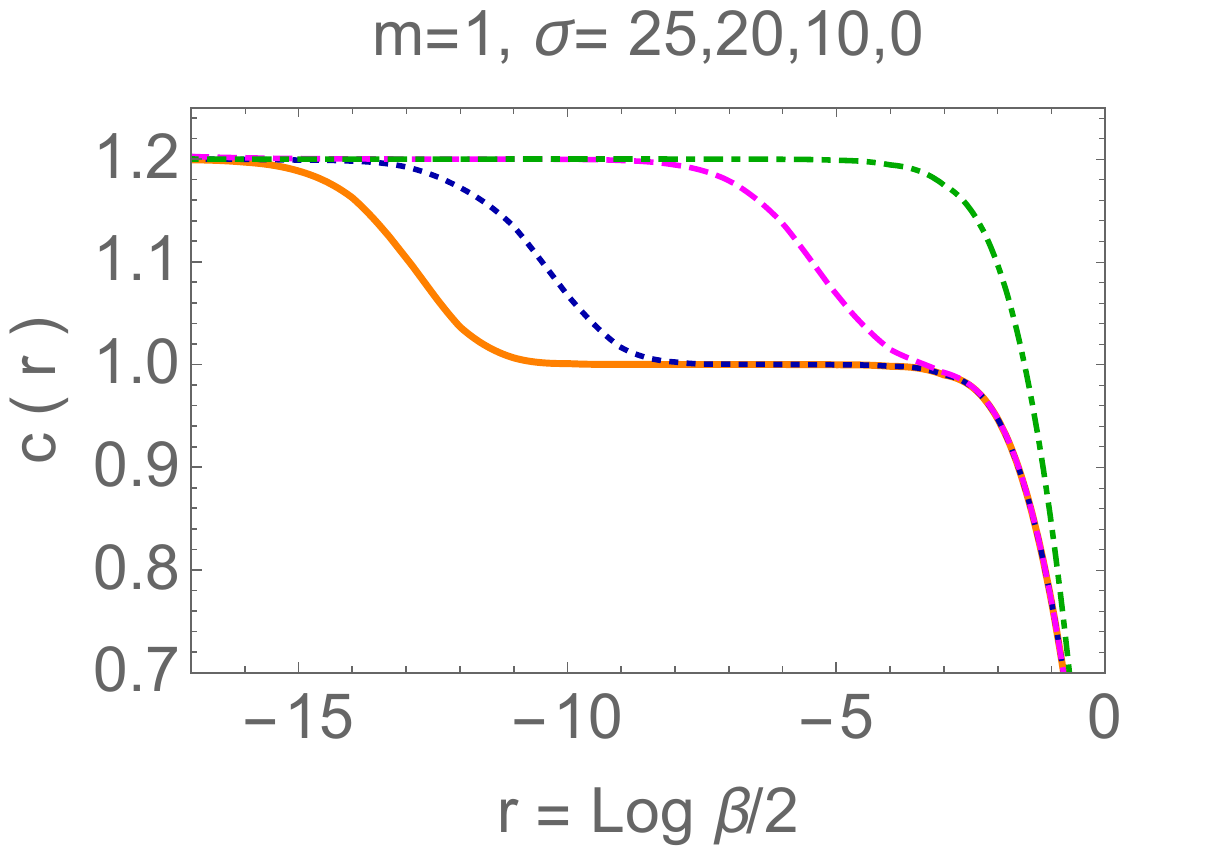} 
 \end{center} 
 \caption{The TBA scaling function of the $SU(3)_2$-HSG model for different values of $\sigma$.}  
 \label{cfun} 
 \end{floatingfigure}}
 A common observed feature of these theories is that many physical quantities, such as those computed in the TBA (e.g.~scaling functions), develop staircase patterns, where the position and size of the steps (or plateaux) are related to the value of $\sigma$. This is in accordance with the separation of energy scales discussed above. Indeed, for large $\sigma$, at temperatures that are large with respect to the mass scale $m$ but small with respect to the separation scale $M$, the theory reaches the UV limit of the two-free-fermion theory, with central charge $c=1$ (as per (\ref{free})). In contrast, for temperatures beyond this separation scale, the UV fixed point is determined by the coset $SU(3)_2/U(1)^2$ and corresponds to $c=\frac{6}{5}=1.2$. The TBA scaling function $c(r)$ with $r=\log\frac{\beta}{2}$ for this model was first presented  in \cite{ourtba}. In Fig~\ref{cfun} we have recalculated it just to give an indication of the structure that we will find for other quantities later on. The various curves correspond to different values of $\sigma$ with the onset of the highest plateau located around $-\frac{\sigma}{2}$.

Thus, when studying RG-flows of different quantities and intermediate values of $\sigma$ we observe that the flows approach these two fixed points in succession, giving rise to the staircase pattern that is typical of this model. From the RG viewpoint, this pattern reflects the presence of a larger amount of degrees of freedom as energy is increased, interaction is turned on, and the unstable particle is formed. As we will see, staircase patterns also emerge in our hydrodynamic analysis, in and out of equilibrium. It is worth noting that staircase patterns in RG flows are also found for other theories, typically the roaming trajectory model \cite{roaming,DRoaming1, DRoaming2} and generalizations thereof \cite{staircase}. However a direct connection to unstable excitations is missing in those cases. A GHD study of the roaming trajectory model with particular emphasis on new scaling functions was recently carried out in \cite{Horvth2019}.

\subsection{Thermodynamic Bethe Ansatz for Generalized Gibbs Ensembles}
\label{tbaeq}

The TBA equations, either at equilibrium in a Gibbs ensemble, or in a GGE, are very simple. The most intrinsic way of fixing the state is through a function, which we will denote $w(\theta;\pm)$, which determines the weight of states in the ensemble. It is such that every state formed of rapidities $\{\theta_i\}$ and particle types $\{\ep_i\}$ has weight $\exp[ - \sum_i w(\theta_i;\ep_i)]$. For instance, in a thermal state at inverse temperature $\beta$, one chooses
\beq
	w(\theta;\pm) = m\beta \cosh\theta \qquad \mbox{(thermal state)}\,,
\eeq
where we already focus on our particular model, with particle species labelled by $\pm$.
One expects that the space of all GGEs with good locality properties (the maximal-entropy states) be spanned by an appropriate space of functions $w(\theta;\pm)$. The TBA provides the full thermodynamics at infinite volumes for such a distribution of states specified by $w(\theta;\pm)$. In the model of interest here, one defines the pseudoenergies $\varepsilon(\theta,\pm)$ by the solution to the system of nonlinear integral equations
\beq
\varepsilon(\theta;\pm)=w(\theta;\pm)-\varphi_{\pm\mp}\star L(\theta;\mp)\,,
\label{pseudoen}
\eeq 
where 
\beq
\varphi_{\pm\mp}(\theta)=-i \frac{d}{d\theta} \log S_{\pm \mp}(\theta)=\frac{1}{\cosh(\theta\pm \sigma)} \quad \mathrm{and}\quad L(\theta;\pm):=\ln(1+e^{-\varepsilon(\theta;\pm)})\,,
\label{kernel}
\eeq
and $\star$ represents the convolution of the functions involved,
\beq
a \star b(\theta):= \frac{1}{2\pi} \int_{-\infty}^\infty a(\theta-\theta') b(\theta') d\theta'\,.
\eeq

From these objects, averages of all local operators can in principle be calculated. We will concentrate on densities $q_i(x,t)$ of conserved charges $Q_i = \int dx q_i(x,t)$, and their currents $j_i(x,t)$ satisfying $\p_t q_i(x,t) + \p_x j_i(x,t)=0$. For these, simple expressions exist. Their averages are fully fixed by giving the one-particle eigenvalues of the associated conserved charge, $h_i(\theta,\pm)$. The averages are obtained by using the ``dressed" quantities $h_i^{\mathrm{dr}}(\theta;\pm)$, which solve the linear integral equations
\beqa
h_i^{\mathrm{dr}}(\theta;\pm)= h_i(\theta;\pm)+ \varphi_{\pm \mp} \star (h_i^{\mathrm{dr}}(\lambda;\mp) n(\lambda;\mp))
\label{dress}
\eeqa 
where
\beq
n(\theta;\pm)=\frac{1}{1+e^{\varepsilon(\theta;\pm)}}
\label{ndefini}
\eeq 
is the occupation function associated to particle $\pm$.
Specifically, the GGE averages of local charge densities $\texttt{q}_i=\langle q_i \rangle_{\underline{\beta}}$ and of their associated currents $\texttt{j}_i=\langle j_i \rangle_{\underline{\beta}}$ are expressed as
\beqa 
\texttt{q}_i &=& 
  \sum_{b=\pm} \int_{-\infty}^\infty \frac{d \theta}{2\pi} e(\theta;b) h_i^{\rm dr}(\theta;b) n(\theta;b)\nonumber\\
&=& \sum_{b=\pm} \int_{-\infty}^\infty \frac{d \theta}{2\pi} e^{\rm{dr}}(\theta;b) h_i(\theta;b) n(\theta;b)\,,
\label{q}
\eeqa 
and 
\beqa 
\texttt{j}_i &=& 
\sum_{b=\pm} \int_{-\infty}^\infty \frac{d \theta}{2\pi} p(\theta;b) h_i^{\rm dr}(\theta;b) n(\theta;b)\nonumber\\
&=& \sum_{b=\pm} \int_{-\infty}^\infty \frac{d \theta}{2\pi} p^{\rm{dr}}(\theta;b) h_i(\theta;b) n(\theta;b)\,,
\label{j}
\eeqa 
(recall that $\underline{\beta}$ was the set of generalized inverse temperatures in the GGE).
Above the energy and momentum one-particle eigenvalues are $e(\theta;\pm)=m \cosh \theta$ and $p(\theta;\pm)=m \sinh \theta$. We have also used a symmetry of the equations that allows us to interchange the ``dressing" operation inside the integral and the sum.

There are a number of natural conserved charges available in the model. For instance, the energy and momentum are local conserved charges which are typically studied in QFT. Further, since in integrable models the scattering is elastic, the number of particles is preserved; thus the total number of particles also is a good conserved charge. In relativistic QFT, this is usually not a local conserved charge, but it is expected to be quasi-local, hence a good characteristic of the state. For the total number of particle and energy one-particle eigenvalues, we will use the notation
\beqa
h_0(\theta;\pm)= 1 && \mathrm{(particle \,\, number)}\, \nonumber\\
h_1(\theta;\pm)=e(\theta;\pm)=m \cosh \theta  && \mathrm{(energy)}\,.
\eeqa
Further, as the scattering is diagonal, the number of particles, energy and other charges carried by each individual particle type are also conserved charges themselves, again expected to be quasi-local. These have one-particle eigenvalues that are nonzero only for one sign of the particle type,
\beqa
 h_0^\ep(\theta;\pm)=  \delta_{\ep,\pm} h_0(\theta;\pm)\,, \qquad\qquad h_1^\ep(\theta;\pm)=\delta_{\ep,\pm} h_1(\theta;\pm)\,.
\eeqa
We will use the notation $\texttt{q}_i^{\ep}$ and $\texttt{j}_i^{\ep}$ for the associated average densities and currents, which therefore take the form
\beq\label{qjpm}
\texttt{q}_i^{\pm}= \int_{-\infty}^\infty \frac{d \theta}{2\pi} e^{\rm{dr}}(\theta;\pm) h_i(\theta;\pm) n(\theta;\pm) \quad \mathrm{and} \quad \texttt{j}_i^\pm=  \int_{-\infty}^\infty \frac{d \theta}{2\pi} p^{\rm{dr}}(\theta;\pm) h_i(\theta;\pm) n(\theta;\pm)\,.
\eeq
Note how the particle types are not summed over in these expressions.

Two intermediate functions in these expressions are of particular interest, as they possess a clear physical meaning: these are the spectral density, and the effective velocities (which first appeared in \cite{BEL14}),
\beq
\rho_p(\theta;\pm)=\frac{1}{2\pi}e^{\rm{dr}}(\theta;\pm) n(\theta;\pm)\,\qquad\mbox{and}\quad
v^{\textrm{eff}}(\theta;\pm)=\frac{p^{\rm{dr}}(\theta;\pm) }{e^{\rm{dr}}(\theta;\pm) }\,,
\label{eff}
\eeq
respectively. The spectral density is a conserved quantity, and the spectral density times the effective velocity, its current, as can be obtained by choosing $h_i(\theta;\pm) = \delta(\theta-\alpha)\delta_{\pm,\ep}$ for any $\alpha,\ep$. Specifically, the quantity $\rho_p(\theta;\pm)d\theta dx$ represents the number of particles of type $\pm$ in a phase-space element $d\theta dx$, while $v^{\rm eff}(\theta;\pm) \rho_p(\theta;\pm)d\theta dx$ is the associated current.

In this paper, we will study numerically the average particle and energy densities and currents, as well as the spectral density and the effective velocities.

Before doing so, one can already extract properties of the dynamics of the model from the structure of the kernels (\ref{kernel}) in the above TBA description:
\bi
\item {\bf Parity Breaking:} The interaction kernels (\ref{kernel}) have standard properties, such as a fast decay at large $|\theta|$, characteristic of the local interaction of the model. For instance, the sinh-Gordon kernel at the self-dual point is $2\, \rm{sech}\,\theta$. However, the kernels (\ref{kernel}) are exceptional in that they are such that parity acts non-diagonally on the asymptotic states. That is, TBA quantities are identical under the simultaneous change of signs of rapidities $\theta\rightarrow -\theta$ and particle types $\pm\to\mp$. This is a remnant of the fact that the the scattering phases \eqref{Smatrix} themselves, and the underlying action of the model, break parity.

\item {\bf Scattering:} The kernels are maximal at $\theta=\mp\sigma$, taking values $\varphi_{\pm\mp}(\mp \sigma)=1$, and rapidly decreasing functions away from their maximum (i.e.~${\rm sech}\,\theta$ is strongly peaked around zero). For instance, for $\sigma>0$, this means that $\varphi_{+-}(\theta)$ is maximal for $\theta= -\sigma<0$. Recalling that $\theta = \theta_1-\theta_2$ is the difference of the rapidities of the two incoming particles with types $\ep_1=+$ and $\ep_2=-$, we see that, for $\sigma$ large and positive, the scattering can be nontrivial only in the region $\theta_1<\theta_2$. This, physically, corresponds to a collision where the particle of type $-$ moves towards the right, and that of type $+$ towards the left, in the rest frame. Analysing $\varphi_{-+}(\theta)$, the same conclusion is reached upon exchanging the roles of $\pm$ particles. Thus nontrivial scattering occurs only in one direction,  for $\sigma>0$ being when particle $-$ travels rightwards towards particle $+$ (and the opposite for $\sigma<0$), and it is this scattering that is expected to give rise to the unstable particle. For this reason, the functions of interest have quite different behaviours for $\theta>0$ and $\theta<0$ for $\pm$ particles, with one choice giving the free fermion result and the other what we can term an ``interaction" result.

\item {\bf Separation into right- and left-movers:} As we know from comparison with soliton gases and the flea gas model \cite{DYC}, the value of the kernels can be interpreted as the distances jumped by particles upon collision. Positive kernels give the ``natural" picture, whereby a tagged particle, travelling rightwards (leftwards) and hitting another particle, experiences a jump leftwards (rightwards), by the amount given by the scattering kernel. Thus, from the previous point, we expect that, say for $\sigma>0$, particle $+$ ($-$) is mostly hit from the left (right) and therefore is mostly displaced toward the right (left); its effective velocity will receive a positive (negative) correction, as compared to its group velocity. We may therefore broadly identify particles of type $+$ with right-movers, and of type $-$ with left-movers. This picture becomes exact near the UV fixed points. In particular, in the presence of the unstable particle, we should find a positive (negative) $+$ ($-$) equilibrium particle current; this will be confirmed by our numerics. 
\ei

\subsection{Out-of-equilibrium steady states}
\medskip

From the results of the previous subsection, given a GGE we may obtain charge density and current averages. We simply solve the equations (\ref{pseudoen}) for the pseudoenergies $\varep(\theta;\pm)$, then obtain $n(\theta;\pm)$, and employ this solution to solve the dressing equation for the conserved quantity of interest, and finally evaluate the integrals (\ref{q}) and (\ref{j}). The main examples we will investigate in this paper are the particle  current  $\texttt{j}_0$, the energy current $\texttt{j}_1$ and the energy density $\texttt{q}_1$, both their total values and the relative contributions  $\texttt{j}^\pm_0, \texttt{j}^\pm_1$ and $\texttt{q}^\pm_1$.

We will now engineer an out-of-equilibrium set up by employing the partitioning protocol, starting with two Gibbs ensembles at inverse temperatures $\beta_{R,L}$ for the right (left) baths. See the reviews \cite{BDReview, VMReview} for the general theory and its applications. As shown in \cite{ourhydro, theirhydro}  hydrodynamic conservation equations and TBA equations can be combined to characterize the steady state currents that emerge in the intermediate region between subsystems at sufficiently large times.

The mathematical procedure goes as follows: equation (\ref{pseudoen}) is solved separately for the right and left sub-systems giving rise to two occupation numbers $n_R(\theta;\pm)$ and $n_L(\theta; \pm)$. One of the main results of \cite{ourhydro,theirhydro} was showing that the non-equilibrium steady state occupation functions occurring at large times at the position $x=0$ are simply
\beq
n(\theta;\pm)=n_R(\theta;\pm) \Theta(\theta-\theta_0^\pm)+n_L(\theta;\pm) \Theta(\theta_0^\pm-\theta),
\label{on}
\eeq
where $\Theta$ is the Heaviside step function. The discontinuity positions $\theta_0^\pm$ are solutions to the equations $p^{\mathrm{dr}}(\theta_0^\pm;\pm)=0$ or, alternatively, they are zeroes of the effective velocities. This form makes a lot of physical sense, as it proclaims that the occupation functions of particles with positive (negative) effective velocities take the form of those in the original ensembles on the left (right) sub-system. Here we assume that the effective velocities are monotonic functions of rapidities, which is confirmed by our numerics below. Therefore, it is easy to carry out a numerical evaluation of the non-equilibrium densities and currents. We give a more detailed description of our algorithm in Appendix B.

Finally, it is worth recalling that for conformal field theory (CFT) the values of $\texttt{j}_1$ and $\texttt{q}_1$ are well-known in this quench protocol. They (and their associated fluctuation spectrum) were investigated in a series of works \cite{BD1,BD2,BD3} and found to be 
\beq
\texttt{j}_1\stackrel{\rm{CFT}}=\frac{c \, \pi}{12} \,(\,T_L^{2} - T_R^{2} \,) \, \quad  \text{and} \quad \texttt{q}_1\stackrel{\rm{CFT}}=\frac{c \, \pi}{12} \,(T_L^{2} + T_R^{2} )\,,
\label{CFT}
\eeq
where $c$ is the central charge and as usual $T_{R,L}=\beta_{R,L}^{-1}$.

\section{Equilibrium Dynamics with Unstable Particles}
We start the main part of the paper by analyzing the equilibrium dynamics of the model. Although the equilibrium properties of this model have been studied at length using TBA techniques, we find that the new ideas brought by the recently developed hydrodynamic picture shed new light into the main features of the theory, especially the nature of the unstable particle. In addition, understanding the equilibrium case in terms of its underlying hydrodynamic properties will be extremely helpful when interpreting the out-of-equilibrium dynamics. 

Throughout this section  we will take $\sigma=20$ and the mass scale $m=1$.
 It is well known  from standard equilibrium TBA arguments that non-vanishing values of the functions  $L(\theta;\pm)$ and $n(\theta;\pm)$ are strongly localized in the range
\beq
\label{cond}
\log{ \frac{\beta}{2}} < \theta < \log{ \frac{2}{\beta} }\,,
\eeq
as the functions fall off double-exponentially outside this range. 
 This observation plays an important role in the design of the numerical algorithm (see Appendix  B). 
Let us now consider several quantities of interest and finally analyse their mutual relationship.
 
 \subsection{Energy Current and Energy Density}\label{suben}
\begin{figure}[h!]
 \begin{center} 
 \includegraphics[width=7cm]{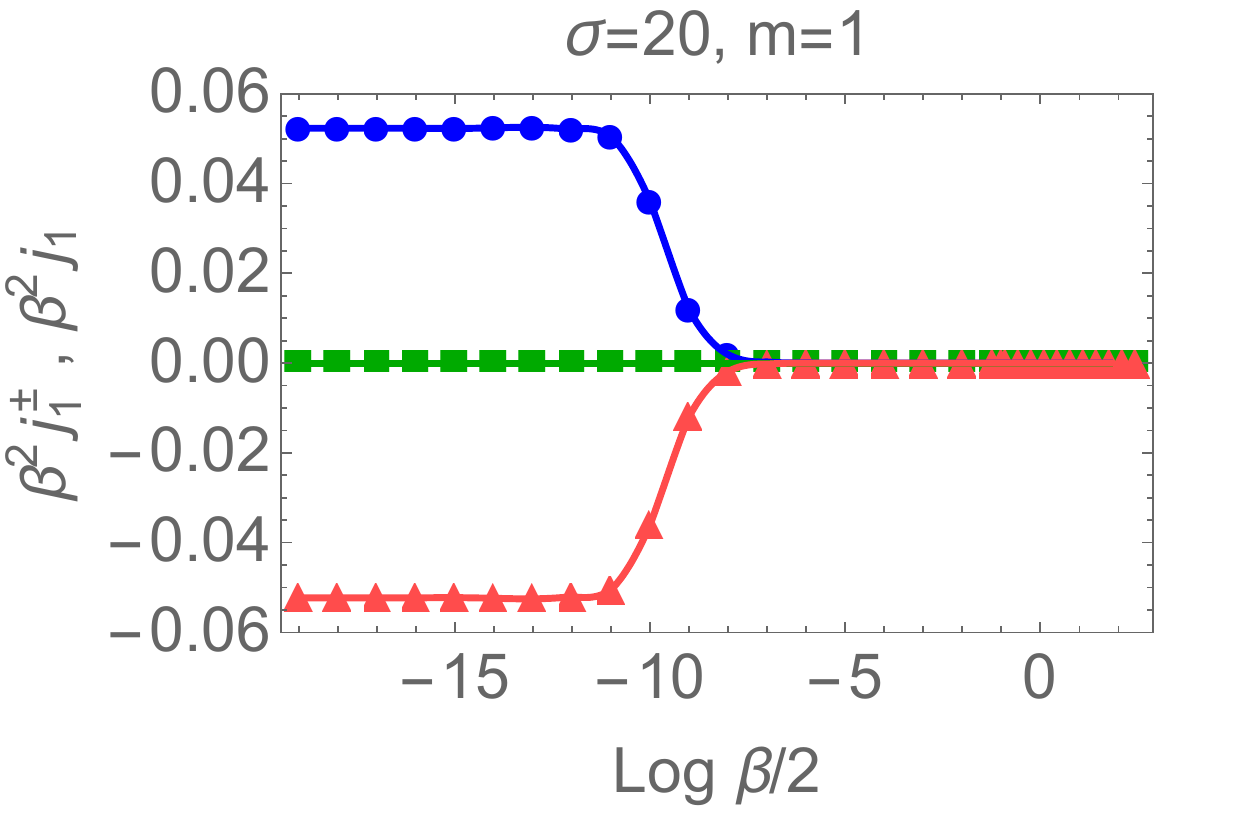} 
  \includegraphics[width=7cm]{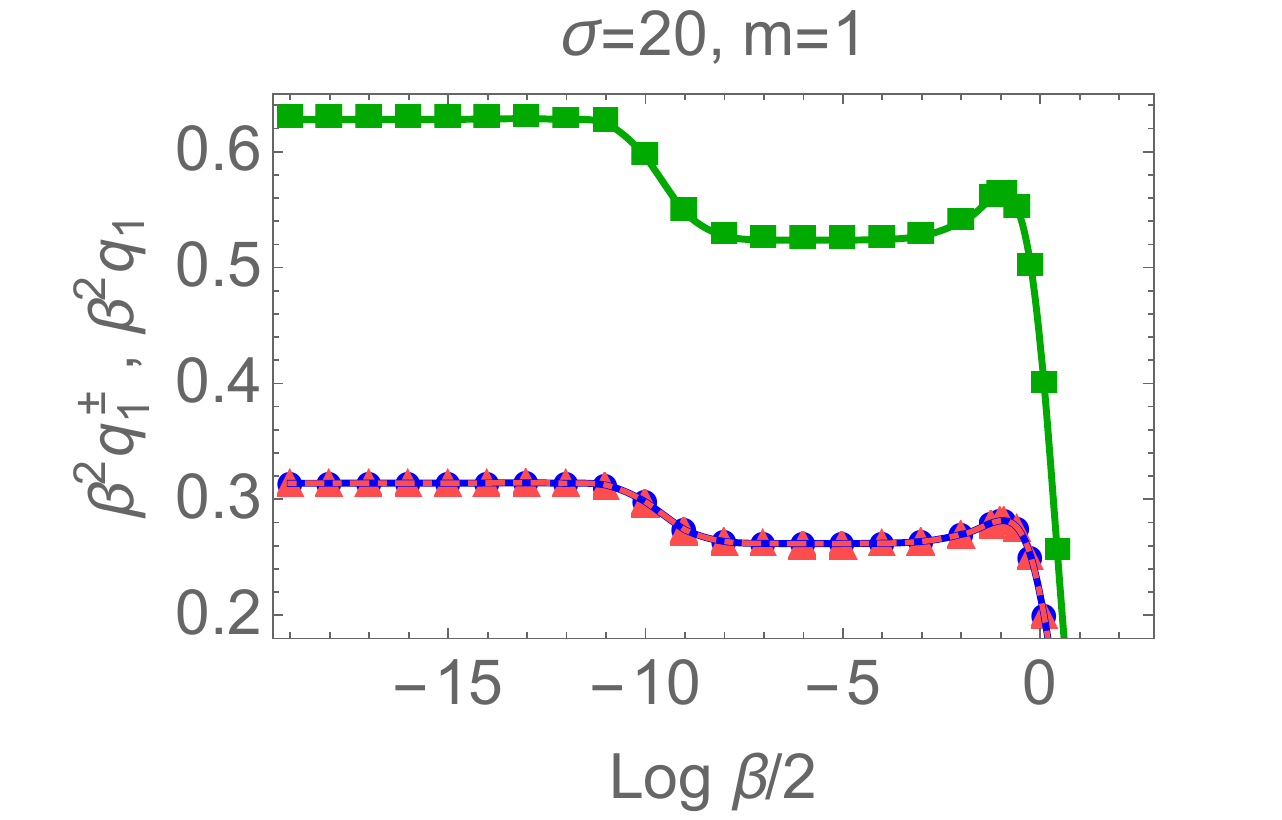} 
 \end{center} 
 \caption{Left: The total scaled energy current $\beta^2 \texttt{j}_1$ (squares, green), the contribution $\beta^2  \texttt{j}_1^+$ (triangles, red) and $\beta^2  \texttt{j}_1^-$ (circles, blue). 
 Right: The total scaled energy density $\beta^2 \texttt{q}_1$ (squares, green), the contribution $\beta^2  \texttt{q}_1^+$ (triangles, red) and $\beta^2  \texttt{q}_1^-$ (circles, blue). In both cases $\sigma=20$ and $m=1$.} 
   \label{equalT}
\end{figure}

  One of the most effective ways to visualize the effect of the unstable particle is to look at temperature-dependent quantities, for a wide range of temperatures.  Evaluating the formulae (\ref{j}), (\ref{q}) and \eqref{qjpm} for the energy ($i=1$), and scaling them by a factor $\beta^2$, we obtain the results of Fig.~\ref{equalT}.  Multiplication by $\beta^2$ is dictated by the CFT result \eqref{CFT}, and is a convenient way to reveal a staircase pattern which reflects the presence of two UV fixed points (with central charges $c=1$ and $c=1.2$), reached for (relatively) low and high temperatures as previously described.

The main properties observed in Fig.~\ref{equalT} are the following:
\bi
\item {\bf Parity Breaking:} Although the total energy current is zero at equilibrium (as expected), the individual contributions from $\pm$ particles are non-vanishing (and opposite) for some energy scales. This is allowed due parity breaking in the theory. More precisely, in TBA, under parity, the signs of the currents and the particle types are exchanged. Here we observe that this gives rise to a negative (positive) energy current carried by $+$ ($-$) particles. 

\item {\bf Onset of the Unstable Particle:} The individual  particle contributions to the energy density and current, and also the total energy density, display a staircase pattern with a step whose onset is located around $\log\frac{\beta}{2}=-\frac{\sigma}{2}=-10$ . This energy value represents the onset of the unstable particle. For $\log\frac{\beta}{2}> -\frac{\sigma}{2}$ the individual contributions to the current are vanishing as this is the regime where the theory behaves as two decoupled free fermions and parity is restored. Energetically speaking, this is the region where energy is not high enough to allow for the formation of the unstable excitation. 

\item {\bf CFT Values:} The staircase patterns observed for the individual contributions to the energy density are identical, because parity preserves the sign of the energy.  Their two plateaux can be predicted from CFT. For lower temperatures $\log\frac{\beta}{2}> -\frac{\sigma}{2}$ the energy densities tend to their massless free fermion value,
\beq
\beta^2 \texttt{q}_1^\pm \stackrel{{\rm{FF}}}=\lim_{\beta \rightarrow 0}\frac{\beta^2}{2\pi}\int_{-\infty}^{\infty} d\theta \frac{\cosh^2 \theta}{1+ e^{\beta \cosh\theta}}= \frac{1}{\pi} \int_0^\infty \frac{u}{1+e^u}\, du=\frac{\pi}{12}=0.261799\,.
\eeq
This corresponds exactly to the height of the lowest plateau of the lower curve on the right panel of Fig.~1.  Similarly, the highest plateau is located at the value 
\beq
\beta^2 \texttt{q}_1^\pm \stackrel{\rm{CFT}}=\frac{\pi c}{12}=0.314159\,,
\eeq
which is the CFT result for $c=1.2$. 
\item {\bf Sign of the Energy Currents:} An interesting feature of Fig.~\ref{equalT} is that the energy current of $+$ ($-$) particles is negative (positive) for high temperatures. However, the structure of the kernels discussed in subsection \ref{tbaeq} suggests that the particle currents should have the exact opposite signs. The solution to this apparent puzzle is that although most $+$ particles propagate towards the right (positive particle current), there are more highly energetic particles that propagate towards the left. The sign of the energy currents of individual particle types is therefore a consequence of an interplay between two phenomena. We fully explain this feature below in our analysis of the spectral densities and effective velocities. 
\ei
\subsection{Effective Velocities}
\label{subeff}
\begin{figure}[h!]
 \begin{center} 
\includegraphics[width=16.5cm]{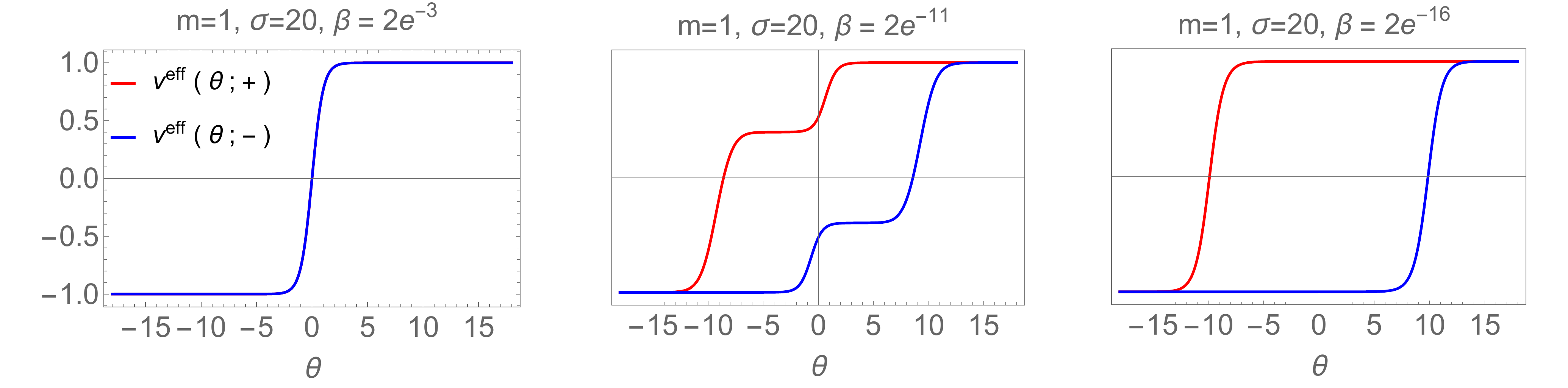} 
\end{center} 
 \caption{Effective velocity profiles at equilibrium for three temperatures: for low temperatures (left figure) we observe the free fermion result, the group velocity $\tanh\theta$; for intermediate temperatures (middle figure) we observe the onset of the unstable particle; for high temperatures (right figure), where a new CFT is reached, effective velocities of $+$ ($-$) particles are shifted so that they appear to be mostly right-moving (left-moving). The evolution of the effective velocities as functions of temperature can be further explored in \href{https://youtu.be/lvWd4qxMShQ}{this video} \cite{video1}.} 
 \label{effvequi}
\end{figure}

 Another interesting quantity to consider are the effective velocities of propagation of the stable particles.  Fig.~\ref{effvequi} shows three ``snapshots" of the velocities as functions of the rapidity variable for three values of the temperature.  
 The main noteworthy features are the following:
\bi
\item {\bf Free Fermion Regime:} For low temperatures  the two velocities are well described by the free fermion result $v^{\mathrm{eff}}(\theta;\pm)=\tanh\theta$. In particular, at large rapidities, we have non-interacting right- and left-movers propagating at the speed of light. 
 \item{\bf Unstable Particle and Parity Breaking:} For intermediate temperatures $\log\frac{\beta}{2}<-\frac{\sigma}{2}$ the onset of the unstable particles triggers a parity breaking effect. Velocity profiles exhibit the symmetry $v^{\mathrm{eff}}(\theta;+)=-v^{\mathrm{eff}}(-\theta;-)$. The presence of the unstable particle marks the presence of interaction and this reduces the absolute values of the velocities, down from their conformal values $\pm 1$. The heights of the intermediate plateaux for both particle types change with temperature until reaching again the values $\pm 1$ at very high temperatures. Some of the features may be explained using the flea gas picture, as explained below.
 
\item {\bf UV Limit:} In the deep UV limit (i.e. very high temperature compared to the unstable particle's mass) the velocities reach once more their CFT values $\pm 1$ but are ``shifted" in comparison with their free fermion value.  In fact they are very well approximated by the functions (\ref{veffap}) which are derived below. We have again large regions of right- and left- movers propagating at the speed of light, and we observe that the $+$ ($-$) particle acquires ``mostly" right-moving (left-moving) properties. This is again in agreement with the flea-gas picture, which, as we explained, indicates that $+$ ($-$) particles should be right-movers (left-movers).

\item {\bf Plateaux and the ``Flea Gas" Picture:} The flea gas scattering picture described at the end of subsection \ref{tbaeq} explains the presence of the intermediate plateaux in the middle panel of Fig.~\ref{effvequi}. For instance, the $+$ particle may only scatter by collisions on its left, and these collisions generate jumps rightwards. Thus, only for $\theta<0$, where the particles are not moving rightwards at the speed of light, can such collisions happen; and when they happen, they ``slow down" the particle. This only happens in a small interval of values of $\theta$ (for the $+$ particle this is approximately the interval $[-\sigma/2,0]$) and the precise boundaries of this intermediate plateau, are more subtle to explain. They are determined by an interplay between spectral densities and the effective velocity. For instance, a change of the effective velocities at rapidities $|\theta|>\sigma/2$ is precluded for low temperatures $\log \frac\beta2 > - \sigma/2$, because no particles are present at such rapidities. The configuration achieved at large temperatures, for instance the right-most panel of Fig. \ref{effvequi}, has however a clear meaning. Indeed, scattering may only happen between $+$ and $-$ particles for rapidity differences near to $\sigma$, but does not happen if particles are co-moving (have the same effective velocity). Thus, for instance, $+$ particles at rapidities $-15$ and $-$ particles at rapidities $5$ do not scatter according to the right-panel of Fig \ref{effvequi}.

\item{\bf Vanishing Velocities:} Interestingly, for $\log\frac{\beta}{2} \approx -10$, that is, precisely at the onset of the unstable particle, the intermediate plateaux both have heights zero. The physical interpretation is that for such temperatures, $+$ and $-$ particles of rapidities $|\theta|<\sigma/2$ are essentially stationary, and this allows them to form the finitely-lived bound state represented by the unstable particle. We will observe the formation of the unstable particle more precisely in subsection \ref{EDIS}.

\ei
The behaviour of the effective velocities for very high temperatures as described in the item on ``UV limit" can be  analytically derived from the TBA equations under
some simple assumptions. Recall the definition of the effective velocities (\ref{eff}) and of the dressing operation (\ref{dress}). We know that the kernels $\varphi_{\pm \mp}(\theta)$ are functions that are strongly peaked around $\theta= \mp \sigma$ and we also know that the functions $n(\theta;\pm)$ develop a plateau in the region (\ref{cond}). For high temperatures this will be a very wide plateau of height $n=\frac{\sqrt{5}-1}{2}=0.618...$ (this can be derived from the constant TBA equations \cite{ourtba}) so that within the region where the kernel is non-vanishing the occupation numbers are constant and may be taken out of the integral. Thus, at high temperatures we can approximately write
\beq
h_i^{\mathrm{dr}}(\theta;\pm)\approx h_i(\theta;\pm)+ \frac{n}{2\pi} \int_{-\infty} ^\infty d\lambda \, \varphi_{\pm \mp}(\theta-\lambda) h_i^{\mathrm{dr}}(\lambda;\mp)\,.
\eeq 
An even cruder approximation consists of treating the kernel as a $\delta$-function $\delta(\theta-\lambda\pm \sigma)$ and writing
\beq
h_i^{\mathrm{dr}}(\theta;\pm)\approx h_i(\theta;\pm)+ n \, h_i^{\mathrm{dr}}(\theta\pm \sigma;\mp)\,.
\eeq 
Assuming that $h_i(\theta;+)=h_i(\theta;-):=h_i(\theta)$ the equations above are solved by the following functions
\beq
h_i^{\mathrm{dr}}(\theta;\pm)=\frac{h_i(\theta)+n\, h_i(\theta\pm\sigma)}{1-n^2} \,.
\eeq
For the effective velocities this means that
\beq
v^{\rm{eff}}(\theta;\pm)\approx \frac{\sinh \theta+n \sinh(\theta\pm \sigma)}{\cosh\theta+ n \cosh(\theta\pm \sigma)}\, \qquad \mathrm{for}\qquad \log\frac{\beta}{2}\gg -\frac{\sigma}{2}\,.
\label{veffap}
\eeq
If $n=1$ the functions above are exactly $\tanh\left(\theta\pm\frac{\sigma}{2}\right)$. In this case $n$ is not $1$ but the function above still resembles a shifted hyperbolic tangent very much. That is the reason why the curves in the rightmost panel in Fig.~\ref{effvequi} look a lot like shifted versions of those in the leftmost panel.

\subsection{Spectral Densities} 
\label{subden}
\begin{figure}[h!]
 \begin{center} 
 \includegraphics[width=5.4cm]{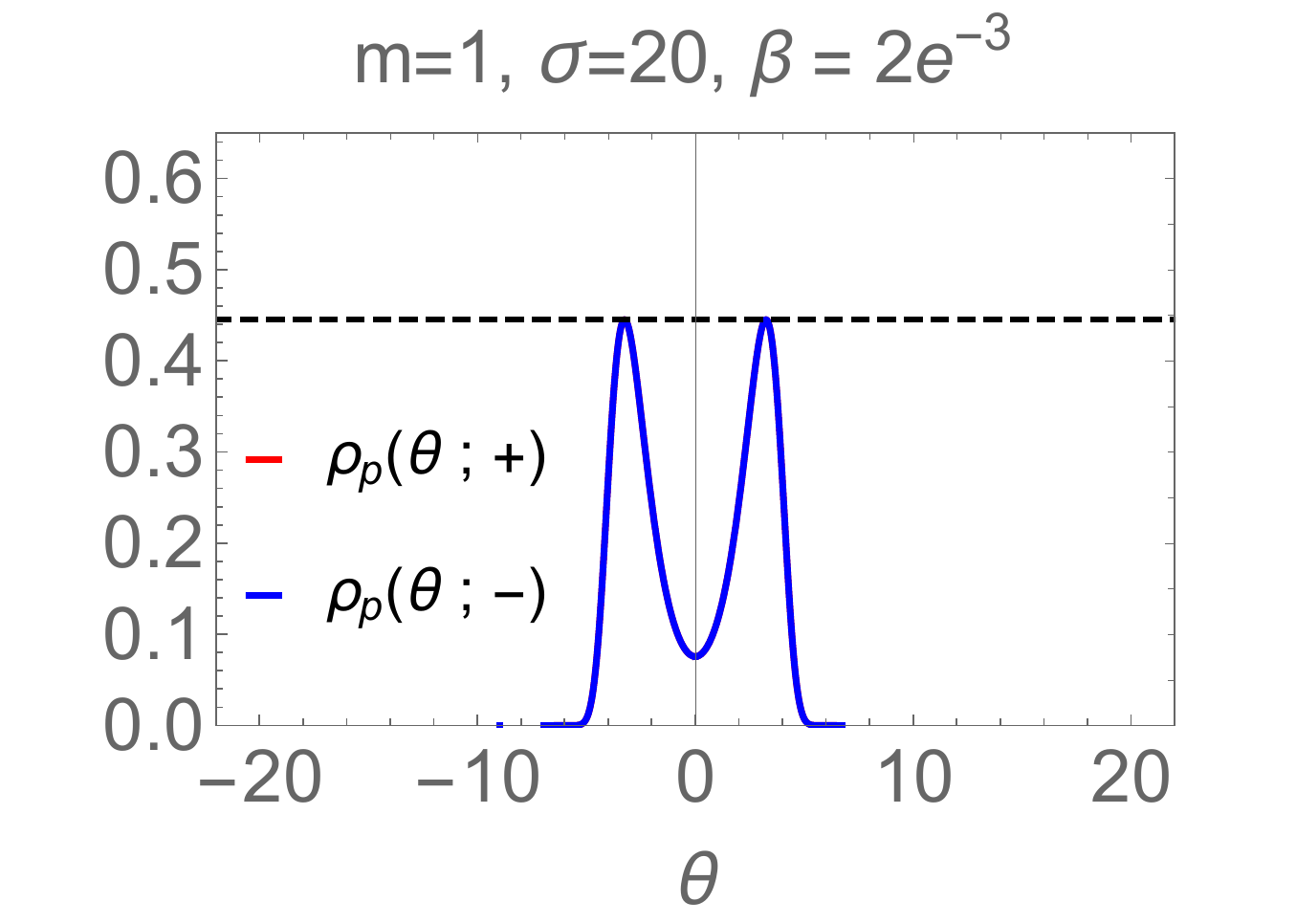} 
 \hspace{-0.4cm}
  \includegraphics[width=5.4cm]{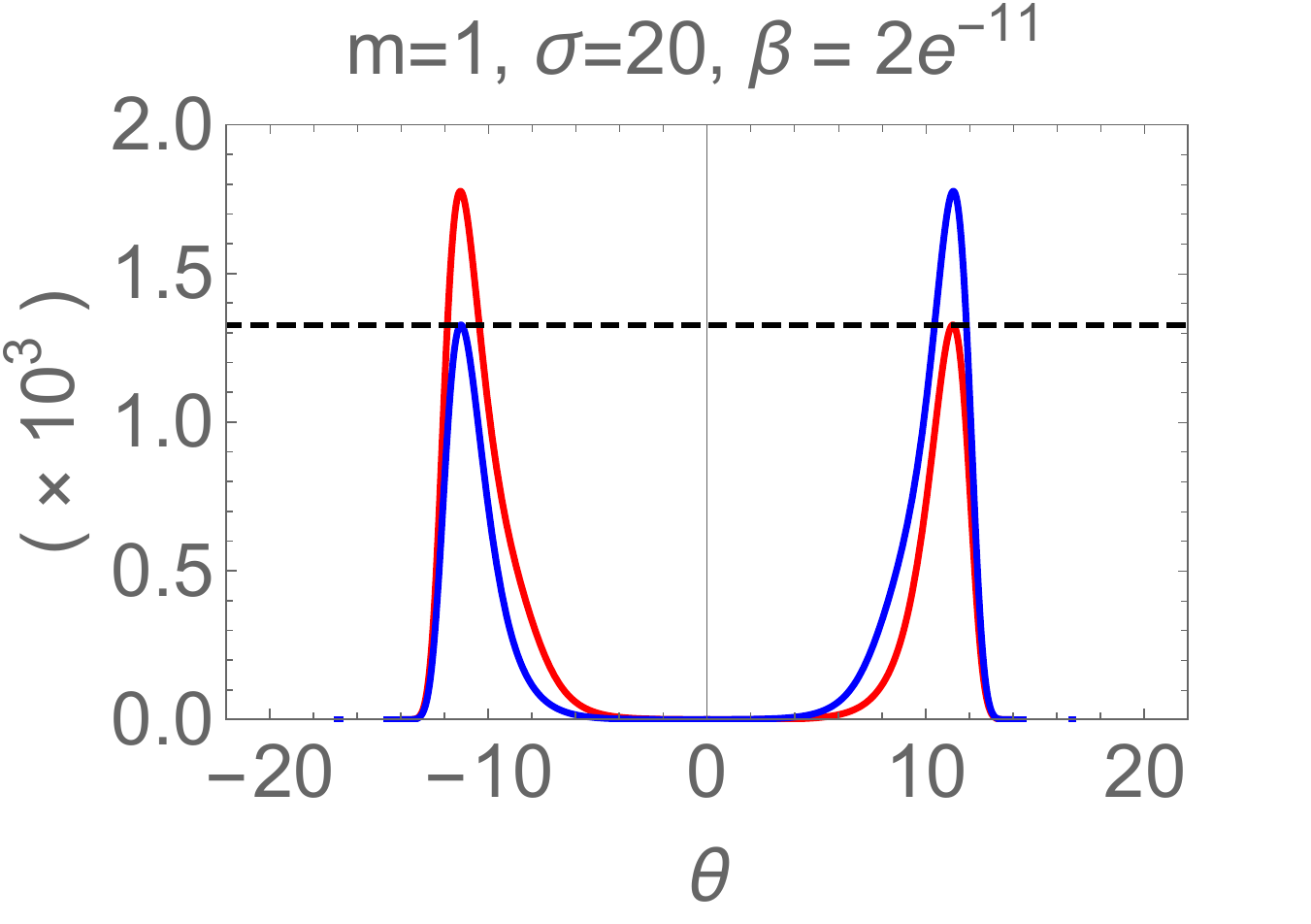}
   \hspace{-0.4cm}
  \includegraphics[width=5.4cm]{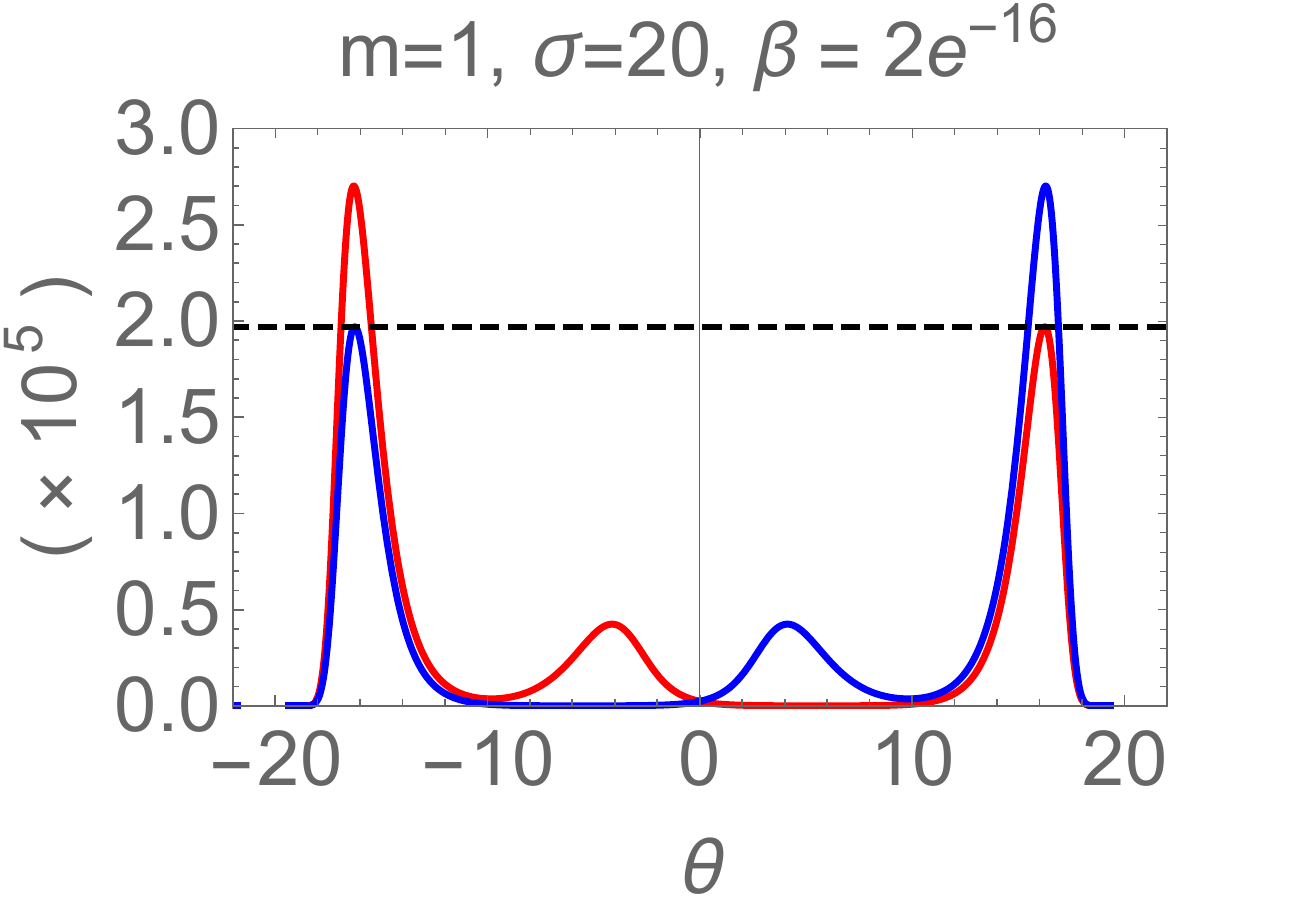}
    \end{center} 
 \caption{Spectral densities for three values of the temperature: $\beta=2 e^{-3}, 2 e^{-11}$ and $2 e^{-16}$. In the two rightmost panels, the vertical axis labels should be multiplied by factors $10^3$ and $10^5$, respectively, as indicated. In all panels, a dashed horizontal line indicates the height of the free-fermion peak, $0.04431.../\beta$. For low temperatures (left panel) we observe the free fermion result; for intermediate temperatures (middle panel) we observe the onset of the unstable particle with one of the peaks growing beyond the free fermion value; for high temperatures (right panel) the densities develop one additional local maximum. The evolution of the densities as functions of temperature can be further explored in \href{https://youtu.be/7jX0HFa1cgs}{this video} \cite{video2}.} 
 \label{denprof}
\end{figure}

In this section we analyse the main features of the spectral densities $\rho_{p}(\theta;\pm)$ defined in (\ref{eff}) by considering three density profiles for low, intermediate and high temperatures. These are presented in Fig.~\ref{denprof}, where, for comparison, the values of the maxima of the free-fermion densities at large temperatues, $\rho_{\rm max}^{\rm FF} = \frac{\ell-1}{2\pi \beta}$ with $\ell=1.27846...$ (dashed black line), are shown.
The most important features of the spectral densities at equilibrium are the following:
\bi
\item {\bf Free Fermion Regime:} The spectral densities for sufficiently low energies (Fig.~\ref{denprof}, left panel) are those of a free fermion and are described by the corresponding formula 
\beq
\rho_p^{\rm{FF}}(\theta;\pm)=\frac1{2\pi} \frac{\cosh\theta}{1+e^{\beta \cosh\theta}}\,.
\label{ffparden}
\eeq
This function has maxima around $\theta=\pm \log\frac{\beta}{2}$, as seen in the figure; more precisely, the positions of the maxima scale, for $\beta$ small, as $\theta \sim \pm \log \frac\beta {2\ell} + o(1)$ where $\ell=1.27846...$ solves $e^{-\ell} = \ell-1$. These maxima are at a height that scales as $\sim \frac{\ell-1}{2\pi\beta}$, as also seen in the figure.
\item {\bf Turning on the Interaction:} For higher temperatures (Fig.~\ref{denprof}, middle and right panels) we still have maxima around $\pm \log \frac{\beta}{2}$, but the heights of some of the maxima start to change as soon as the unstable particle comes into play. For intermediate energies we observe that for each given particle type, one of the maxima (the right (left) one for $+$ ($-$) particles) coincides with its free fermion value whereas the other maximum is higher, indicating an ``excess" density generated by the onset of the interaction. This asymmetry is justified by the structure of the kernels, as discussed in subsection~\ref{tbaeq}. That is the $\varphi_{+-}(\theta)$ kernel is maximized at $\theta=-\sigma<0$ and is negligible for $\theta>0$ thus the effect of interaction only manifests itself for $\theta<0$ while  the free fermion physics persists for $\theta>0$.
\item {\bf Three Local Maxima:} For high temperatures (compared to the unstable particle's mass)  two new local maxima, one for each density, emerge located around $\pm (\log \frac{\beta}{2}+\sigma)$ (Fig.~\ref{denprof}, right panel). Thus, at high temperatures, each spectral density exhibits three local maxima: the free fermion peak expected for that temperature, the ``interacting peak" whose maximum is largest, and a smaller, ``subsidiary peak". We observe two important features for these peaks. First, the position of the maxima is once more justified by the scattering matrix which dictates that interaction is maximized for rapidity differences $\pm \sigma$. In particular, the rapidity difference between the $+$ particle (red) interacting peak and the $-$ particle (blue) subsidiary peak is, at all temperatures, around $-\sigma$, the value at which the scattering interaction $\varphi_{+-}(\theta)$ is maximal; and viceversa. Second, for each particle type, the excess area of the interacting peak compared to the free fermion peak roughly coincides with the area of the subsidiary peak. By combining with a dynamical analysis, these features are fully explained in the next subsection.
\ei

\subsection{Scattering, Dynamics and the Unstable Particle}
\label{EDIS}
We now argue that by simultaneously analysing features of the effective velocities and spectral densities, we gain a new, dynamical insight into the equilibrium scattering theory of the model.

The conventional understanding of unstable particles is based on the presence of a pole in the scattering amplitudes and on the notion of how the presence of this particle adds, at large temperatures, new degrees of freedom to the theory: it drives an RG flow between, in the IR, a double free fermion theory and, in the UV, a non-trivial coset model. However, the introduction of dynamical quantities such as the effective velocities, in combination with the two observations we have made in the last point of subsection \ref{subden}, brings a new, perhaps more intuitive perspective into the interpretation of this unstable particle.

\begin{figure}[h!]
\begin{center} 
 \includegraphics[width=16cm]{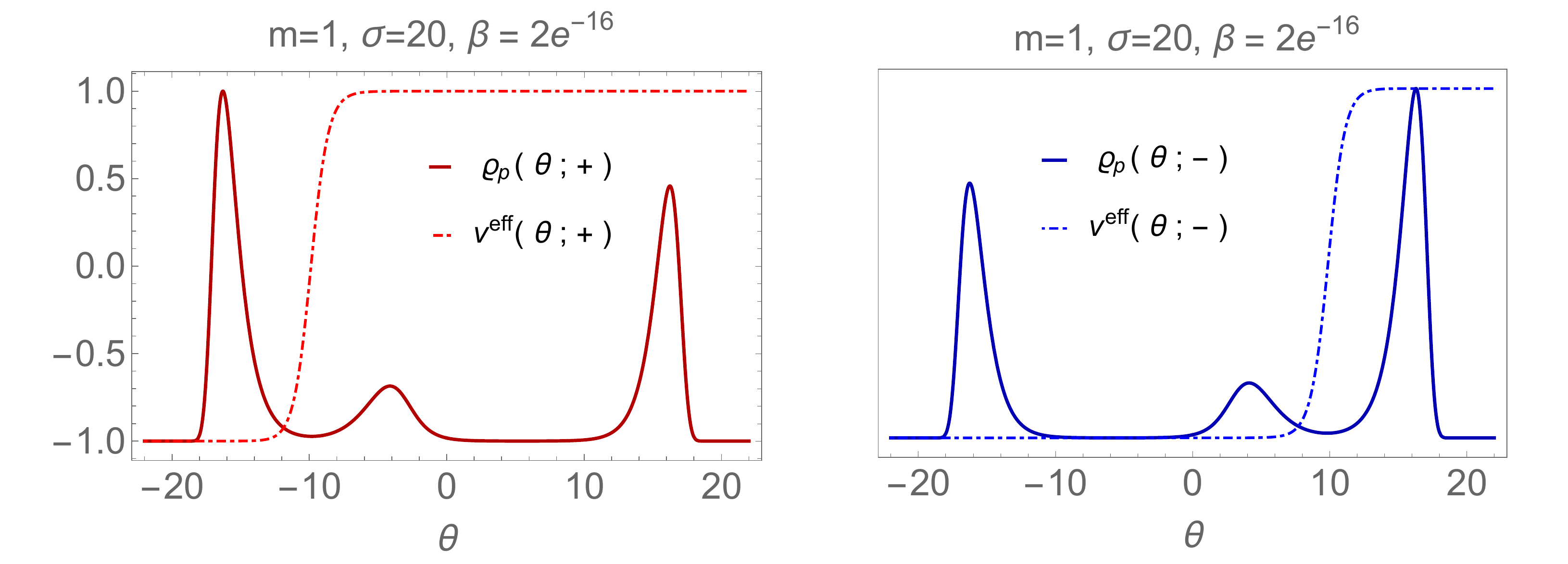} 
    \end{center} 
    \caption{The effective velocities versus the normalized spectral densities $\varrho_p ( \theta ; \pm )= 2 \rho_p ( \theta ; \pm ) / \tilde{\rho}^{ \, \pm} -1 $ where $\tilde{\rho}^{ \, \pm}$ is the height of the largest local maximum. The inverse temperature  is $\beta=2 e^{-16}$.}
\label{extra}
\end{figure}

We illustrate this with Fig.~\ref{extra}, which shows the same high temperature physics we have seen in subsections \ref{subeff} and \ref{subden} and combines scaled versions of the curves found in the right panels of Figs.~\ref{effvequi} and \ref{denprof}. Consider the positions of the local maxima of the spectral densities in Fig.~\ref{extra} and the corresponding values of the velocities. For particle $+$ (left panel, red) the density has maxima around $\log\frac{\beta}{2}\approx+ 16$ (free fermion peak), $\approx -4$ (subsidiary peak) and $\approx -16$ (interacting peak). Comparing with the effective velocity curve, the particles these represent have velocities very nearly $+1$, $+1$ and $-1$, respectively. For particle $-$ (right panel, blue), the maxima of the free fermion, interacting and subsidiary peaks are around $\log\frac{\beta}{2}\approx - 16$, $\approx 4$ and $\approx +16$, respectively, with velocities $-1$, $-1$ and $1$, respectively.

Thus, the velocities associated with the interacting peak of each distribution and the subsidiary peak of the other distribution are {\em always the same}. These particles are co-moving, staying parallel to each other for all times, and thus have the opportunity to bond. Since, as we observed in subsection \ref{subden}, their rapidity separation $\pm\sigma$ are at the maxima of the scattering kernel $\varphi_{\pm\mp}(\theta)$, these particles are indeed subject to a strong interaction, and can form bound states (even if only finitely-lived). Further, as the ``excess" area of the interacting peaks are roughly the same as the areas of the subsidiary peaks, the excess density created by the onset of interaction and the subsidiary peak can be interpreted as pairs of bound $(+-)$ and $(-+)$ particles propagating at the same speed. These are the unstable particles, gathered within two clouds, one right-moving and one left-moving.  The population of unstable, finitely-lived particles thus formed is rendered stable by the high energy of the thermal bath and the continuous availability of co-moving, interacting particles of opposite types.

In summary, varying the temperature and observing the various structures form with their respective effective velocities, is  the most direct way we know of ``visualizing" the formation of the unstable particle. This visualization is particularly striking when observing the continuous change of the densities as temperature is increased in \href{https://youtu.be/7jX0HFa1cgs}{this video}.

Counting the degrees of freedom, over all, we therefore have two free fermions (each with its right- and left-moving components), and, in addition, one unstable particle (also with its right- and left-moving components). These degrees of freedom lead to the central charge $c=1.2$ seen in the UV. In order to account for it quantitatively, we need to look at the energy per unit temperature-square carried by the particles, by multiplying the spectral density by the factor $(12\beta^2/\pi) e(\theta;\pm)$. The total area under the curves is then the central charge -- a measure of the total number of degrees of freedom. This can be also seen as a consequence of the CFT result (\ref{CFT}). We then observe that the contributions coming from the excess area of the interacting peaks largely dominate the contributions from the subsidiary peaks, as the subsidiary peaks are at smaller values (in absolute values) of rapidities. Therefore, the contribution of each particle to the energy current and densities, and to the central charge, coming from the clouds of unstable particles is dominated by the excess density of the interacting peak. This total excess energy area indeed represents $1/5$ of the energy area covered by the free fermion parts, as it should.

Interestingly, for $+$ particles, say, the interacting peak consists of particles propagating with velocity $-1$. As this dominates any contribution from the subsidiary peak, this means that there are more highly energetic $+$ particles propagating with velocity $-1$ than there are with velocity $+1$. Therefore a negative energy current is generated, as observed in subsection \ref{suben}. Similar arguments can be made for the $-$ particles. This is against the naive intuition from the flea-gas picture, which suggests that $+$ ($-$) particles are right- (left-)moving. Here we see that it is due to the energetics of the additional degree of freedom that appears in the UV and the associated propagation velocities. The naive intuition is recovered when looking at the particle currents themselves, instead of the energy currents, as we do in the next subsection.

\subsection{Particle Currents}
\begin{figure}[h!]
 \begin{center} 
 \includegraphics[width=7.5cm]{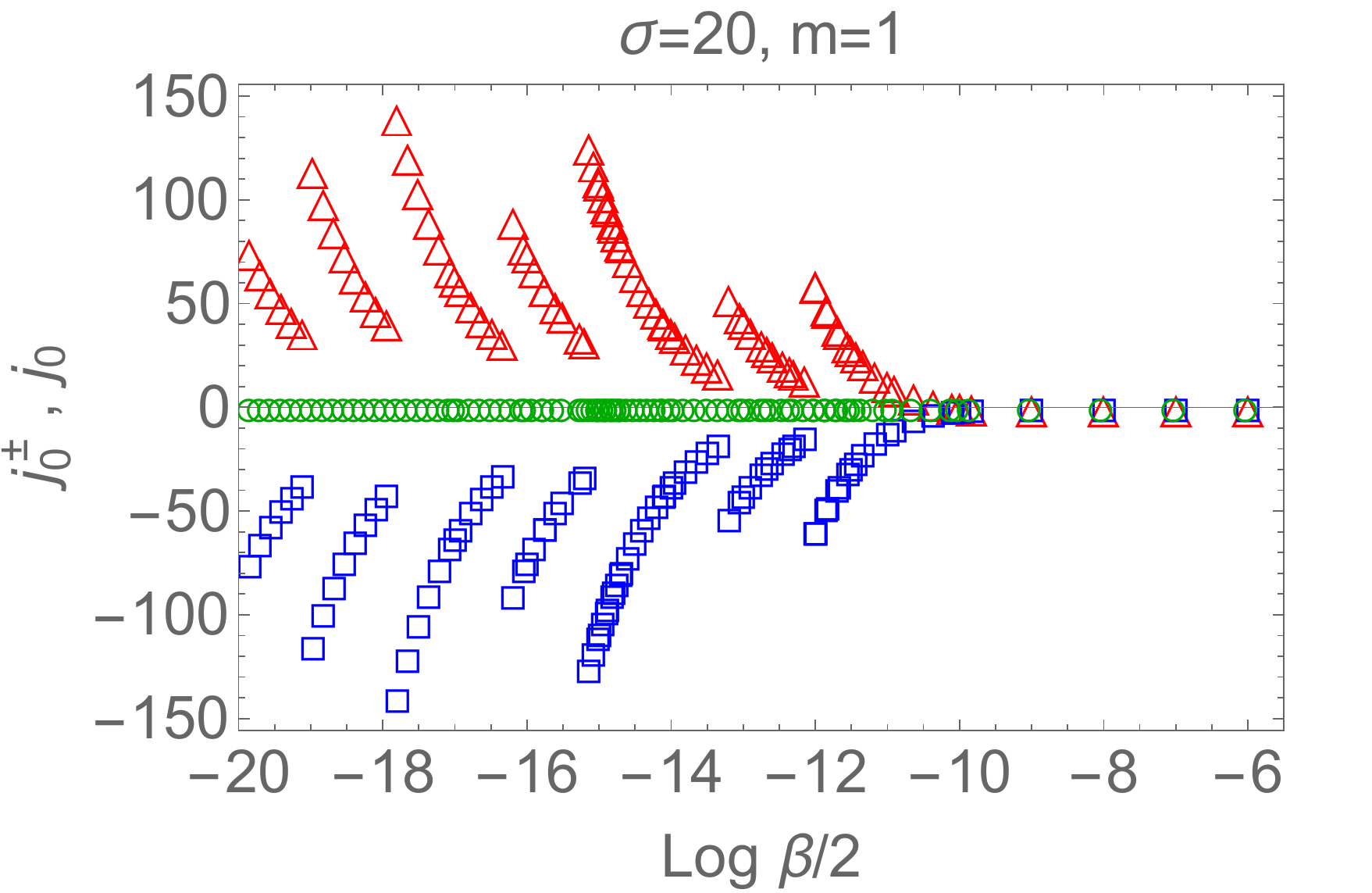} 
  \hspace{-0.2cm}
  \includegraphics[width=7.2cm]{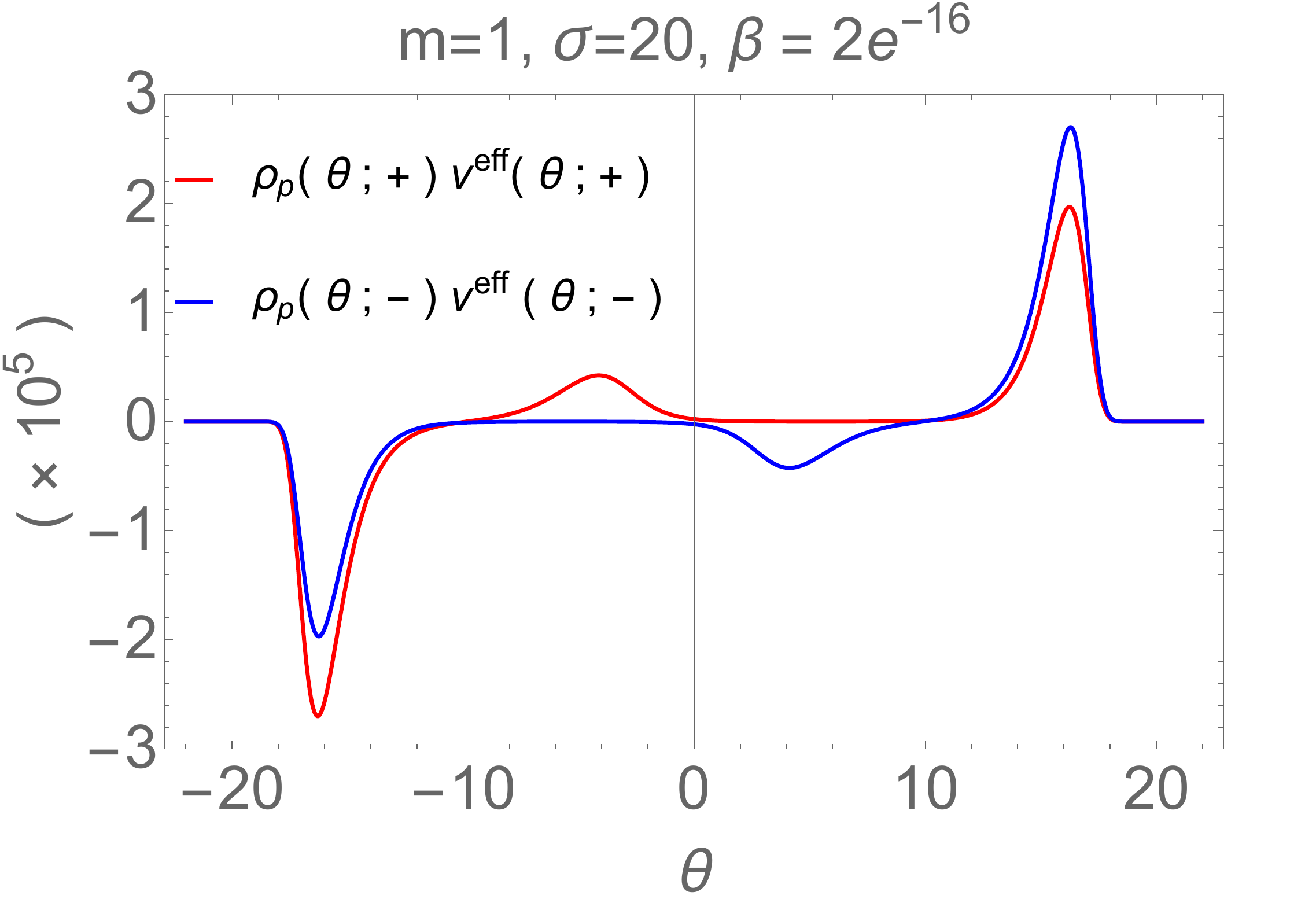}
    \end{center} 
 \caption{Left panel: Particle currents at equilibrium as functions of temperature. Right panel: The functions $\rho_p(\theta;\pm) v^{\rm{eff}}(\theta;\pm)$ for $\beta=2 e^{-16}$ whose integration gives the particle currents. Here the vertical axis should be multiplied by a factor $10^5$.} 
 \label{parcu}
\end{figure}

In this section we take a brief look at the particle currents as functions of temperature. They are presented in Fig.~\ref{parcu}. 
We note the following main features:
\bi
\item {\bf Free Fermion Regime:} As for the energy, the particle currents are zero in the free fermion regime $\log\frac{\beta}{2}>-\frac{\sigma}{2}$. 
\item {\bf Turning on the Interaction:} For $\log\frac{\beta}{2}<-\frac{\sigma}{2}$ when interaction is not negligible anymore, we observe non-zero currents.
 As predicted by the scattering picture of subsection~\ref{tbaeq} the current is positive (negative) for particle $+$ ($-$). 
\item {\bf A Tight Balance:} The particle currents are (relative to other quantities) very small. 
In contrast the functions whose integration they result from (Fig.~\ref{parcu}, right panel)  take very large values (both positive and negative) but positive and negative values are almost perfectly balanced so that in the end only a very small current is produced. This is because, as noticed earlier, the total area of two smaller peaks (free fermion and subsidiary) of spectral density is approximately equal to the area of the larger (interacting) peak. From the velocity profiles this suggests that there are approximately as many particles of each type moving with velocity $\pm 1$.
\item {\bf  An Intricate Structure:} In view of the above we may wonder if the intricate structure observed in Fig.~\ref{parcu}, left panel is a true physical effect or a numerical error. 
We have performed several tests, increasing precision substantially and found the structure is robust. Thus we believe it is an accurate result. However, for the moment we have no plausible physical interpretation for the structure of these functions. 
\ei

\section{Out-of-Equilibrium Dynamics with Unstable Particles}
We will now analyse the same quantities as in the previous section in an out-of-equilibrium situation. As described previously, two thermal baths at inverse temperatures $\beta_L, \beta_R$ are connected at $t=0$. As before our analysis focuses on the case $\sigma=20$ with mass $m=1$. For simplicity we have also chosen the position at the origin, so we look at the ``ray" that is located exactly in the middle of the steady-state region. The physical picture does not change substantially for other rays. In much of our analysis we will fix the ratio of temperatures and vary $\beta_L$ only. We will use the new variable:
\beq
x=\frac{\beta_R}{\beta_L}\,,
\eeq
(not to be confused with the position variable which is never explicitly used in our formulae).
Hence $x>1$ corresponds to a positive temperature gradient $T_L>T_R$ and $x<1$ corresponds to a negative temperature gradient $T_L<T_R$.
\subsection{Energy Currents and Energy Densities}
\begin{figure}[h!]
 \begin{center} 
 \includegraphics[width=7.5cm]{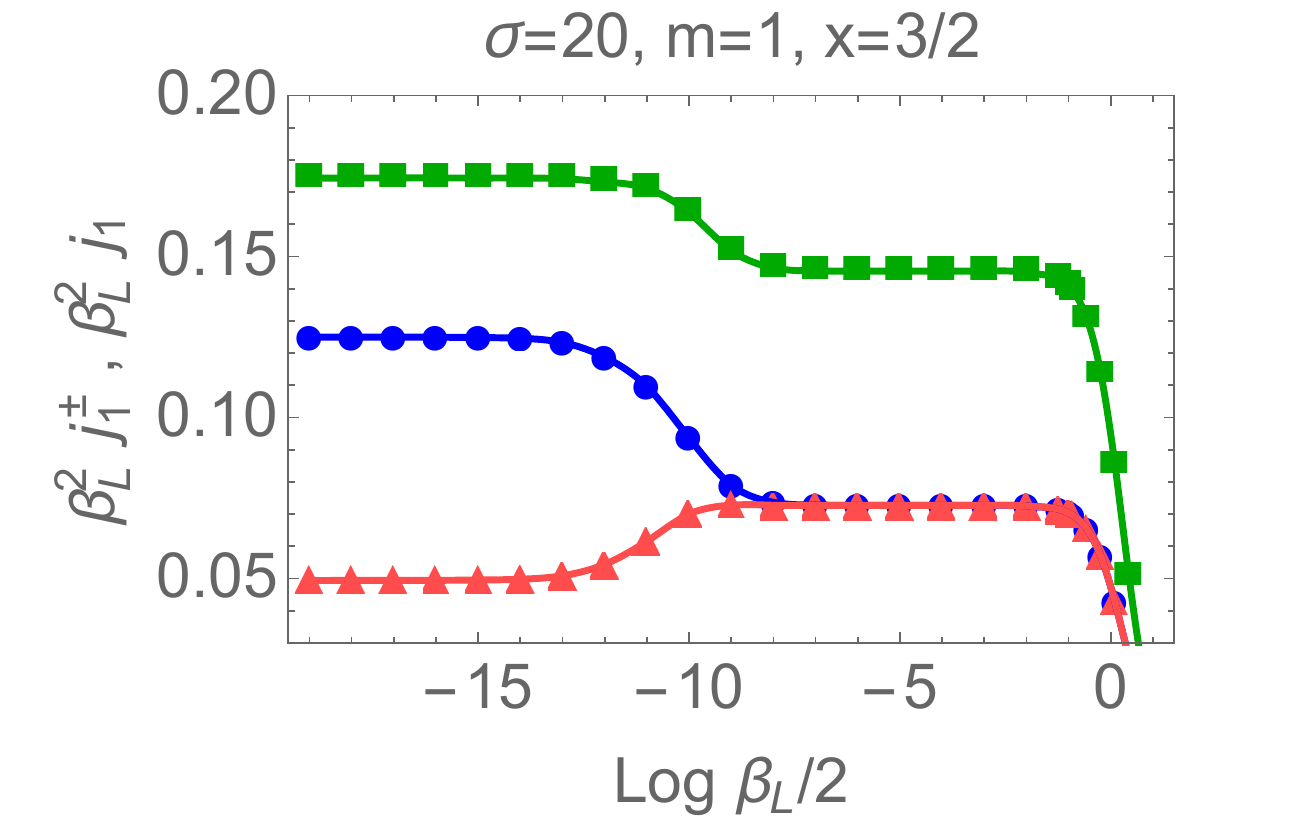} 
  \hspace{-0.15cm} \includegraphics[width=7cm]{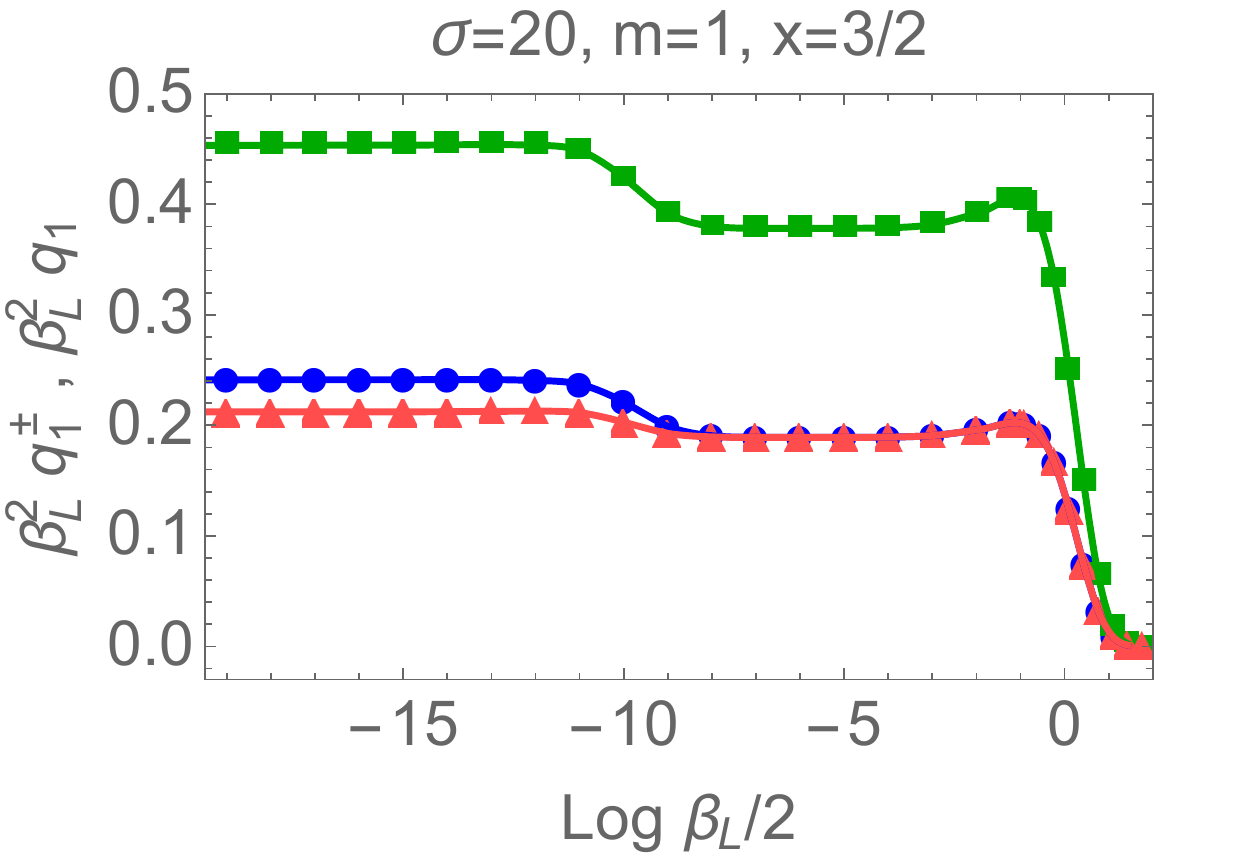}
  \includegraphics[width=7.5cm]{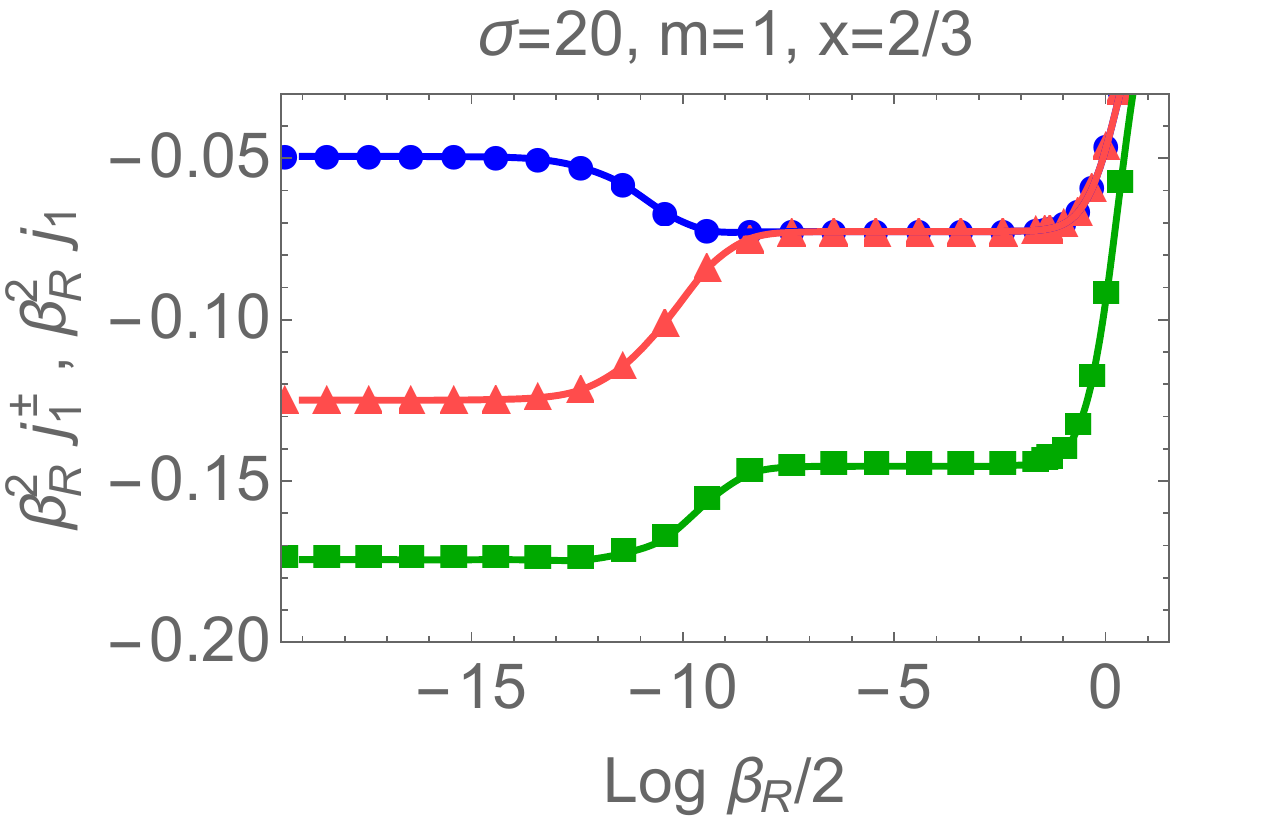} 
   \hspace{-0.1cm} \includegraphics[width=7cm]{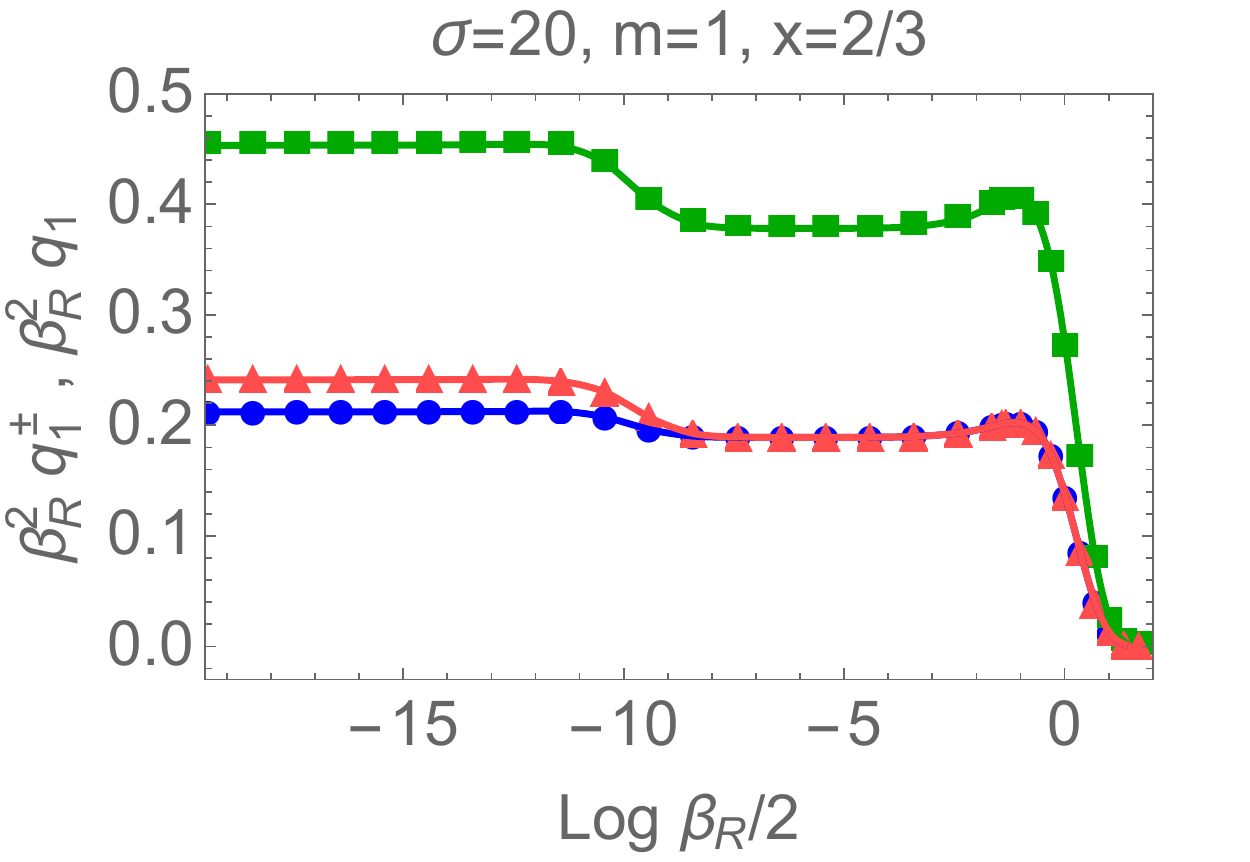}
    \end{center} 
 \caption{The total (scaled) energy current $\beta_i^2 \texttt{j}_1$ (squares, green), the contribution $\beta_i^2  \texttt{j}_1^+$ (triangles, red) and $\beta_i^2  \texttt{j}_1^-$ (circles, blue) and similarly for the energy density. We consider the cases $x=3/2$ ($i=L$) and $x=2/3$ ($i=R$). 
  In all cases $\sigma=20, m=1$.} 
  \label{difT}
\end{figure}

In this section we discuss the main features of the out-of-equilibrium energy currents and energy densities for different temperature ratios, focussing on the main changes with respect to the equilibrium situation. Our discussion focusses on Fig.~\ref{difT}.
The main important features  are the following:
\bi
\item {\bf Symmetry:} A clear feature from the pictures is the following symmetry under the exchange $x \rightarrow x^{-1}$ (or $\beta_L \leftrightarrow \beta_R$):
\beq
 \texttt{j}^\pm_1(\beta_L, \beta_R) \rightarrow - \texttt{j}^\mp_1(\beta_R, \beta_L) \qquad \mathrm{and} \qquad  \texttt{j}_1(\beta_L, \beta_R) \rightarrow - \texttt{j}_1(\beta_R, \beta_L)\,,
\eeq
and similarly
\beq
 \texttt{q}^\pm_1(\beta_L, \beta_R) \rightarrow  \texttt{q}^\mp_1(\beta_R, \beta_L) \qquad \mathrm{and} \qquad  \texttt{q}_1(\beta_L, \beta_R) \rightarrow  \texttt{q}_1(\beta_R, \beta_L)\,.
\eeq
This is in agreement with the parity symmetry of the TBA equations.
\item {\bf Conformal Limits:}
 The height of the plateaux is predicted as in the equilibrium case by the formula (\ref{CFT}). For instance, for $x=3/2$ the scaled total current has plateaux at
\beq
\beta_L^2 \texttt{j}_1\stackrel{\rm{CFT}}=\frac{\pi c}{12} \left(1-\frac{4}{9}\right)=\frac{5 \pi c}{108}\,,
\eeq
which gives values $0.174533$ and $0.145444$, for $c=1.2$ and $c=1$, respectively. The same holds for the total spectral density:
\beq
\beta_L^2 \texttt{q}_1\stackrel{\rm{CFT}}=\frac{\pi c}{12} \left(1+\frac{4}{9}\right)=\frac{13 \pi c}{108}\,,
\eeq
predicting the values $0.453786$ and $0.378155$ for $c=1.2$ and $c=1$.

\item {\bf Unstable Particle Onset:} In all figures we also see the location of the start of the plateau at $-\sigma/2=-10$ with respect to the scales $\log \frac{\beta_{L,R}}{2}$. In fact, quantities associated with particle $+$ develop a plateau for $\log\frac{\beta_L}{2}>-\frac{\sigma}{2}$ whereas for particles of $-$ type the plateau's onset occurs at $\log\frac{\beta_R}{2}=-\frac{\sigma}{2}$. This is hardly detectable in these figures because $\log \frac{3}{2}=0.405...$ and therefore there is little difference between the values $\log \frac{\beta_{L,R}}{2}$; but we have verified this fact for larger values of $x$. 

\item {\bf Particles Couple Mainly to one Bath:} The previous point suggests that type $+$ particles are particularly sensitive to the value of $\beta_L$ whereas particles of type $-$ couple strongly to the value of $\beta_R$. This is related to the structure of the kernels described in subsection \ref{tbaeq} and also to the structure of the occupation numbers (\ref{on}). For particle $+$ this means that it will feel strong interaction with particle $-$ only when $\theta<0$ and close to $-\sigma$. At the same time, for $\theta<0$ the occupation number is largely described by its equilibrium value on the left bath (see Fig.~\ref{theta*} for more details) and so particle $+$ mainly interacts at inverse temperature $\beta_L$. A similar argument can be made for particle $-$.

\item {\bf Equilibrium Currents vs Temperature Gradient:}  In contrast to the equilibrium case, here both particle type contributions to the currents have the same sign, although they are different from each other.  For $x>1$ both contributions are positive, even though the contribution of particle $+$ is always smaller (the opposite is true for $x<1$). This change can be explained as the result of interference (sometimes constructive, sometimes destructive) between two phenomena: the equilibrium dynamics and that induced by the temperature gradient. If $x>1$ we have that $T_L>T_R$ and so from the temperature gradient we expect a positive current. However, for particle $+$ the equilibrium current would have the opposite sign and so, even if temperature ``wins" in the end, we still have a reduced current. For particle $-$ on the other hand both the gradient and the equilibrium tendency support a positive current, so its total contribution is enhanced. The opposite effect is seen for $x<1$.

\begin{figure}[h!]
 \begin{center} 
 \includegraphics[width=7.9cm]{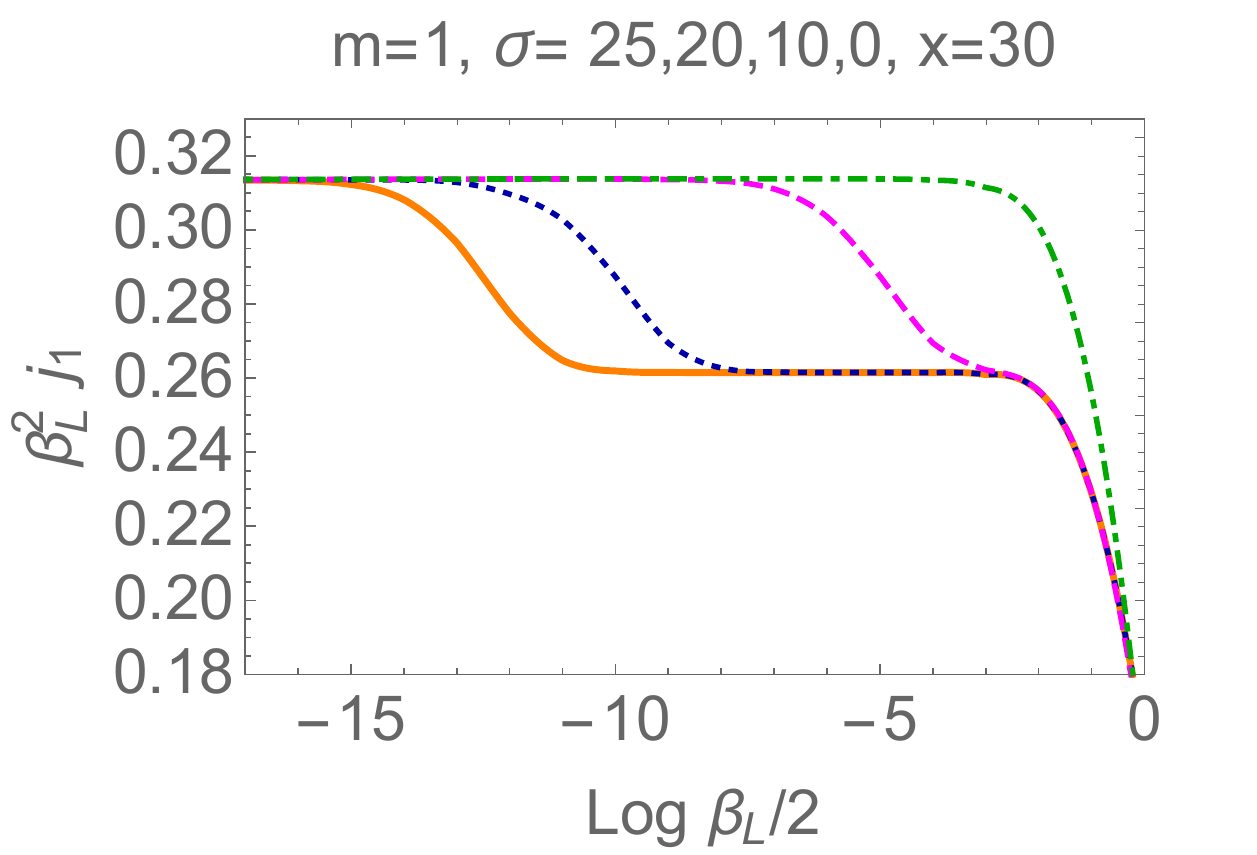} 
      \end{center} 
 \caption{The scaled energy current as a $c$-function for several values of $\sigma$ and temperature ratio $x=30$. The plateaux are located at $\beta_L^2 j_1\stackrel{\rm{CFT}}=\frac{\pi}{12}(1-\frac{1}{900})=0.261508$ and $\beta_L^2 j_1\stackrel{\rm{CFT}}=\frac{\pi}{10}(1-\frac{1}{900})=0.31381$.} 
 \label{enercfun}
\end{figure}

\item{\bf Out-of-Equilibrium $c$-Theorems:} As we have seen, it is possible to read off the central charges of the various UV points that are visited by the theory as temperatures are increased by, for instance, computing the energy current or the energy density. Therefore, one may think of the quantities $ \frac{12 |j_1| \beta^2}{\pi},  \frac{12 q_1 \beta^2}{\pi}$, where $\beta$ is the largest temperature, as new scaling functions. They are qualitatively similar to two well-known distinct scaling functions: the standard TBA function $c(r)$ depicted in Fig.~\ref{cfun} and Zamolodchikov's $c$-function \cite{Zamc}. Several examples exhibiting the same qualitative features as  Fig.~\ref{cfun} are presented in Fig.~\ref{enercfun}.  This idea is however not new. Indeed many such scaling functions were proposed in the work \cite{solong} and, more recently, for the roaming trajectory model in \cite{Horvth2019}.
\ei

 \subsection{Effective Velocities}
 \label{sooeveff}
\begin{figure}[h!]
 \begin{center} 
 \includegraphics[width=7.9cm]{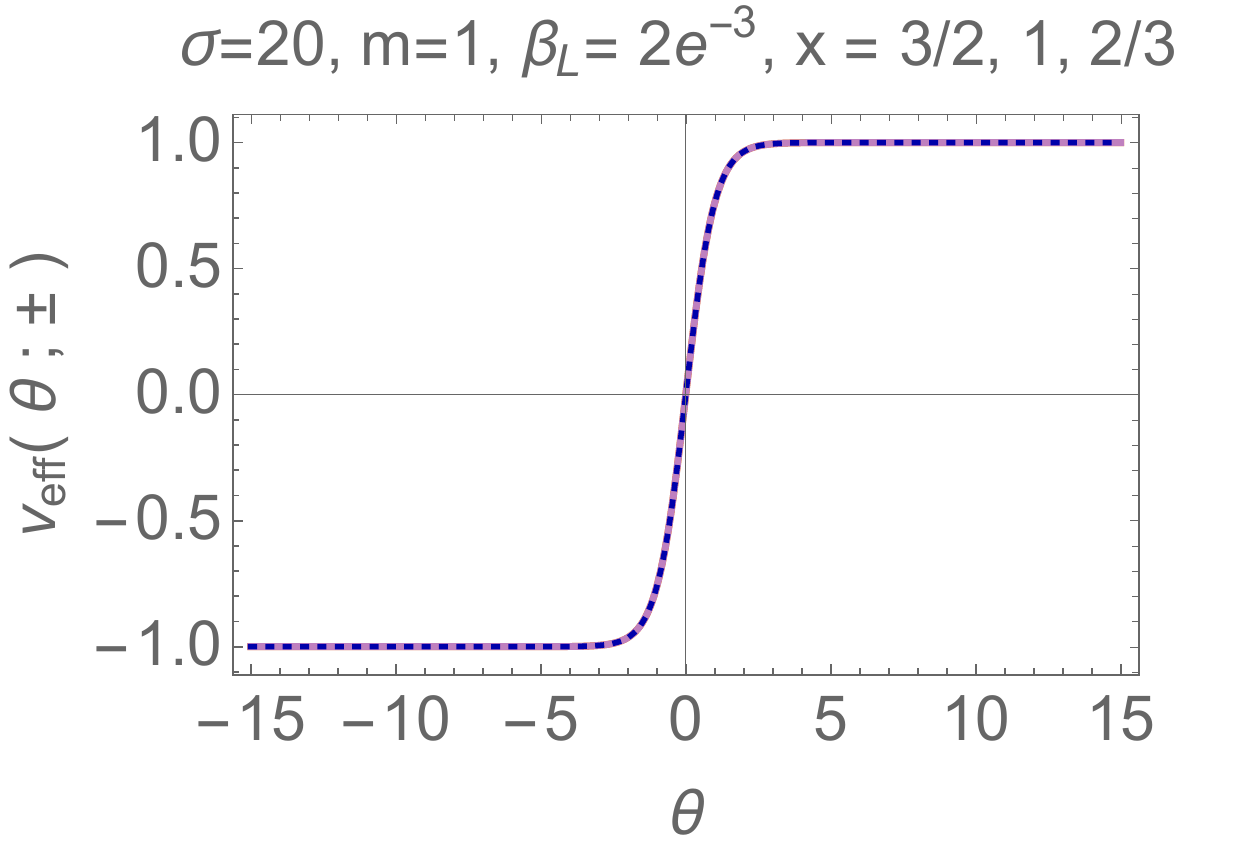} 
   \includegraphics[width=7.9cm]{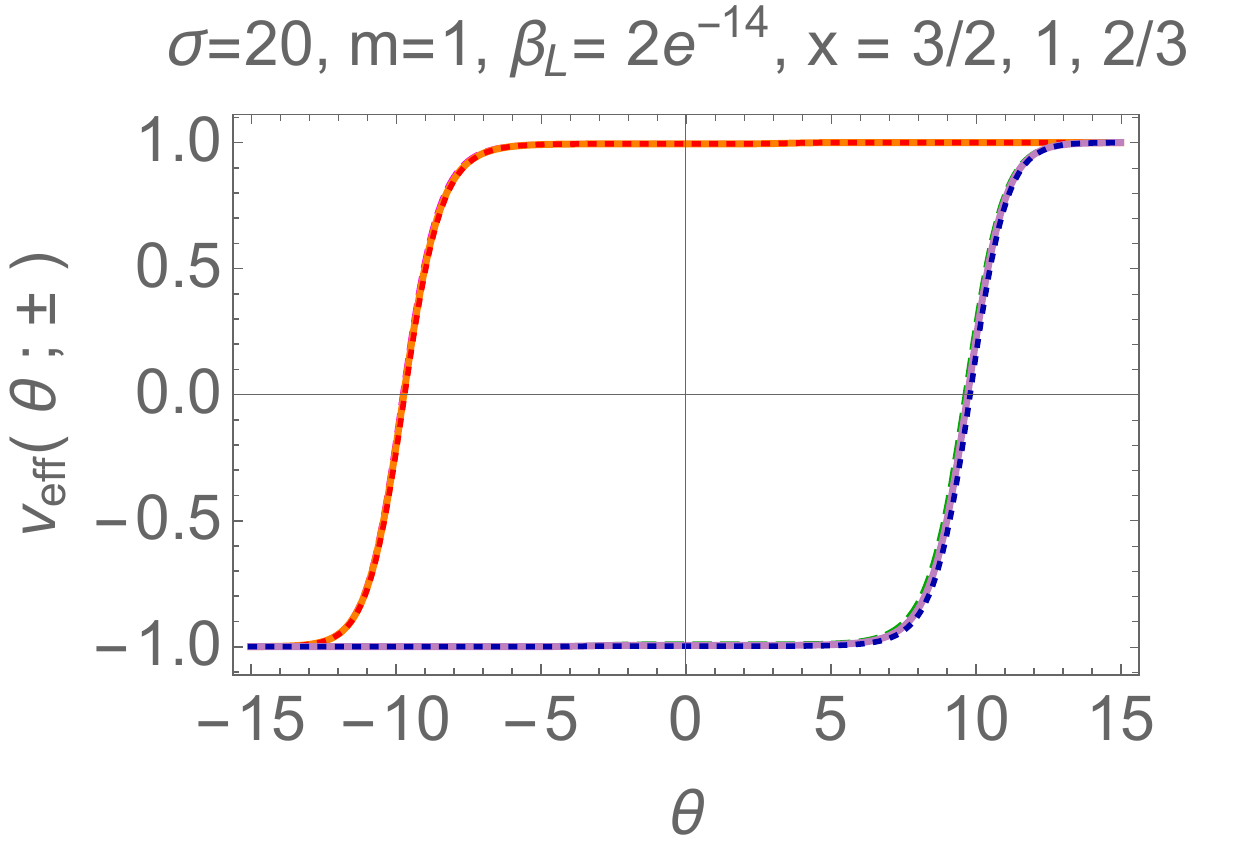} 
 \includegraphics[width=7.9cm]{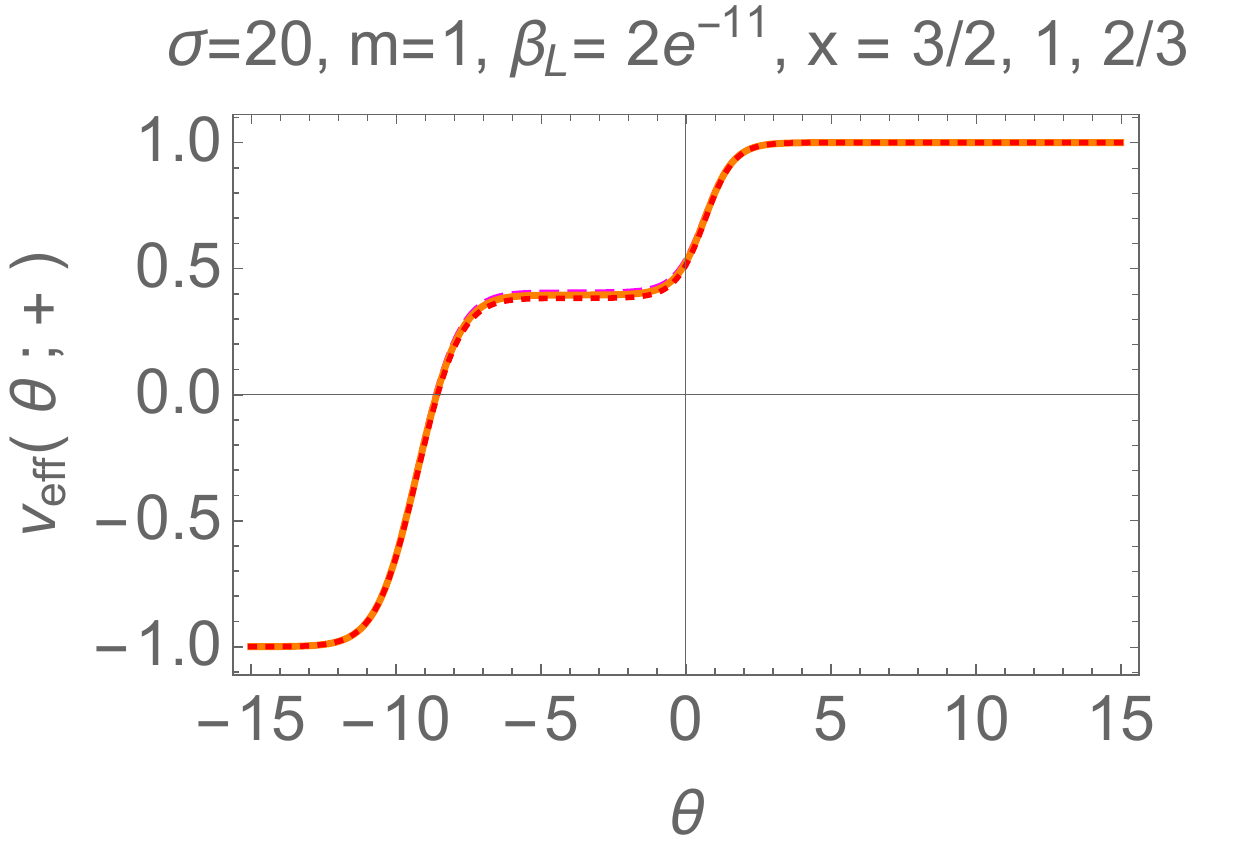} 
  \includegraphics[width=7.9cm]{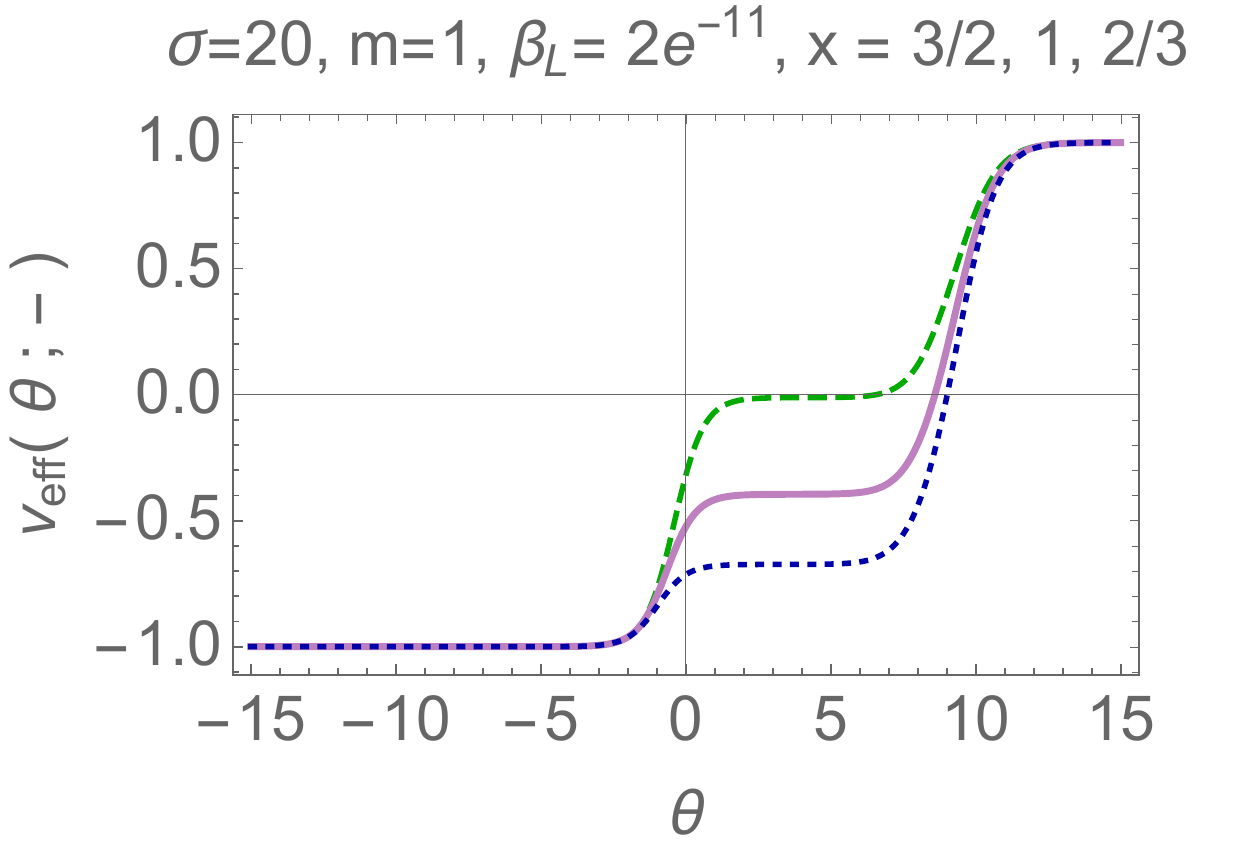} 
     \end{center} 
 \caption{Steady state effective velocities for three values of  $\beta_L$ and three values of $x$, including (for comparison) the equilibrium case $x=1$. The velocity profiles retain many of their equilibrium features. In the bottom right figure $x=\frac{3}{2}$ (dashed, green), $x=1$ (solid, pink) and $x=\frac{2}{3}$ (dotted, blue). The variation of the velocities with temperature can be further explored in 
 \href{https://youtu.be/m2ApWaQkcHE}{this video} \cite{video3}.} 
 \label{ooeveff}
\end{figure}

In this section we take another look at the effective velocities with a focus on changes with respect to the equilibrium behaviour. Fig.~\ref{ooeveff} explores this behaviour for low, intermediate and high temperatures. 
Our main observations are the following:
\bi
\item {\bf Conformal Regime:} Both at low and high temperatures the equilibrium behaviours are recovered. For low temperatures we find the free fermion result. For very high temperatures we find the conformal equilibrium result. Once temperature is high enough the UV result is approached even if $\beta_L\neq \beta_R$.

\item {\bf Velocity Signs:} The velocity of particle $+$ ($-$) is positive (negative) for most rapidities. Except for the free fermion regime, it exhibits a zero for a negative (positive) rapidity value. This is confirmed by Fig.~\ref{theta*} in the appendix. Despite this feature and the fact that for very high temperatures the velocities are identical to their equilibrium values, the particle currents for $x>1$ in this case are both large and positive, indicating that spectral densities are maximal around values of $\theta$ for which both velocities are positive, see subsection \ref{suboutpartcurrent}. 

\item {\bf Particles Couple Mainly to one Bath:} 
For intermediate temperatures, like the ones considered in the second row of Fig.~\ref{ooeveff}, we observe that whereas $v^{\rm{eff}}(\theta;+)$ is virtually unchanged as long as $\beta_L$ is fixed, even if $\beta_R$ is changed, $v^{\rm{eff}}(\theta;-)$ is very much dependent on the values of $\beta_R$. This can be explained by the same arguments presented in the previous subsection. 

\item {\bf Effective Velocities Zeroes:}  The height of the intermediate plateau of the velocities that emerges for intermediate temperatures changes with temperature so that there exists a choice of temperatures
$\log\frac{\beta_{R}}{2} \approx -10$ for which the plateau of the $-$ particle velocity is at height zero (as on the dashed green line in the bottom right panel of Fig.~\ref{ooeveff}) and similarly for particle $+$. This suggests that the effective velocities at this particular temperature have a continuous set of zeroes. The results for $\theta_0^\pm$ shown in Fig.~\ref{theta*} indicate however that although the position of the zero of the effective velocities is extremely sensitive to the temperature around the onset of the unstable particle, it is still a continuous, single-valued function. 
\ei

\subsection{Spectral Densities}
\begin{figure}[h!]
 \begin{center} 
 \includegraphics[width=5.4cm]{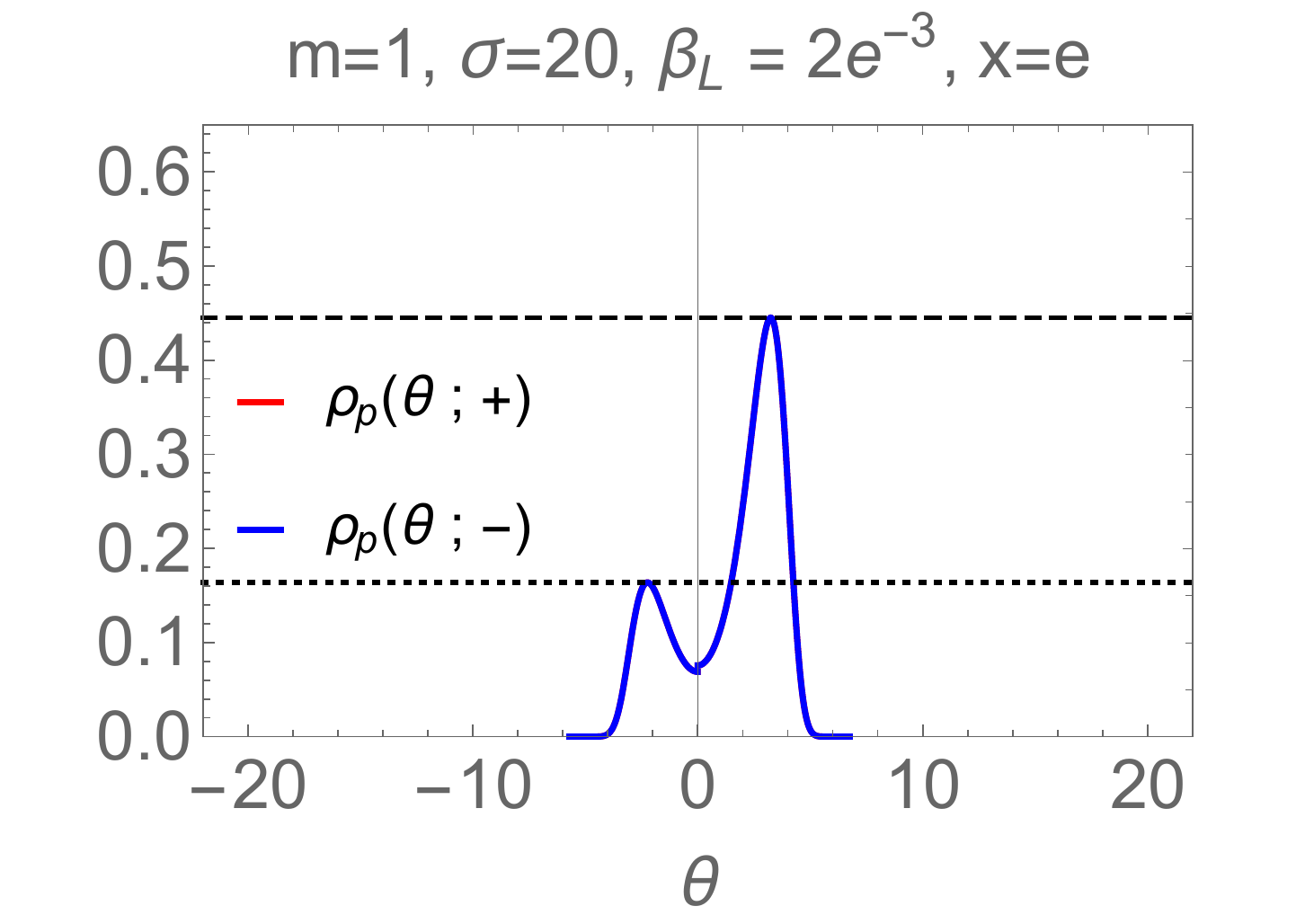} 
 \hspace{-0.4cm}
 \includegraphics[width=5.4cm]{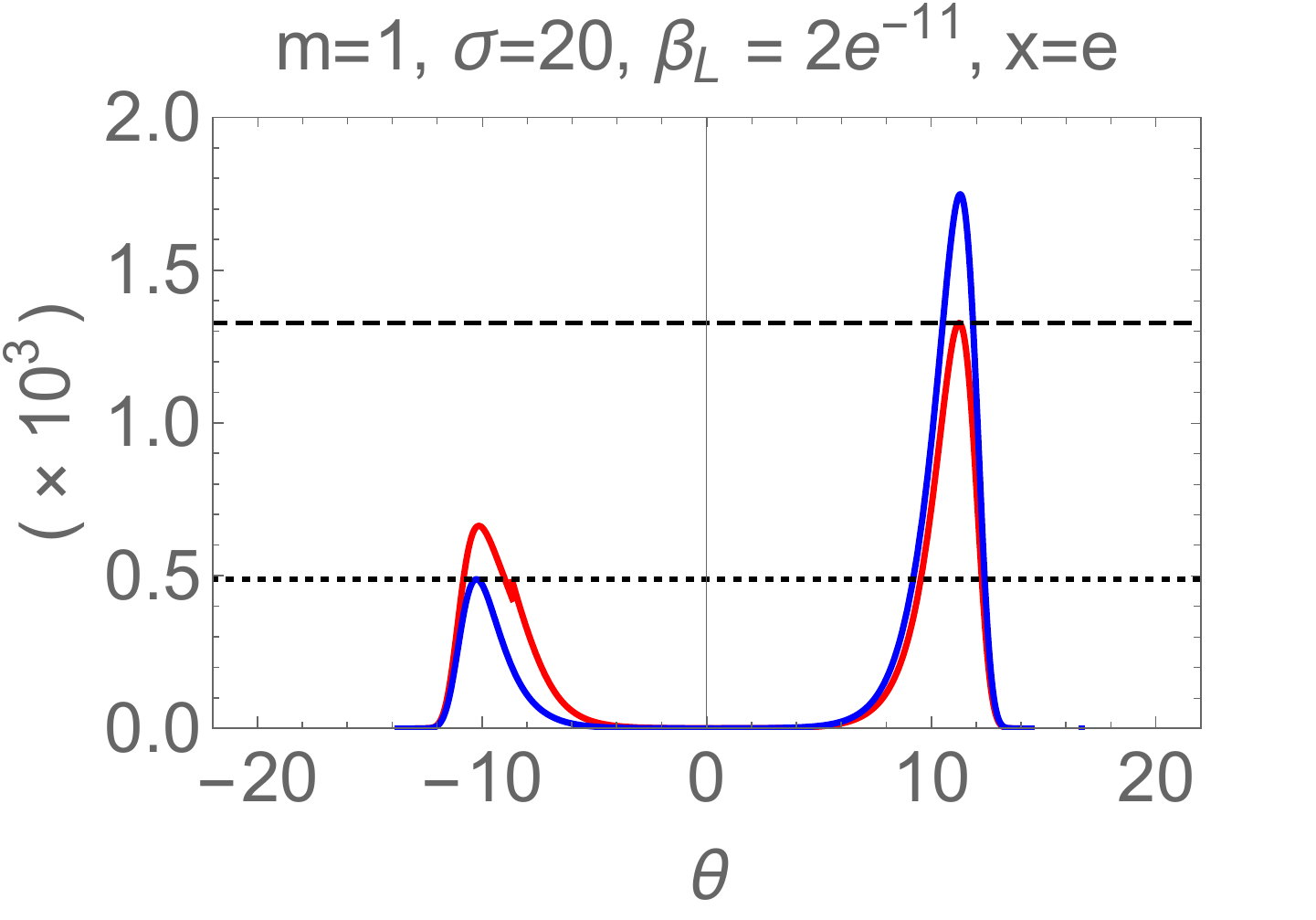} 
  \hspace{-0.4cm}
  \includegraphics[width=5.4cm]{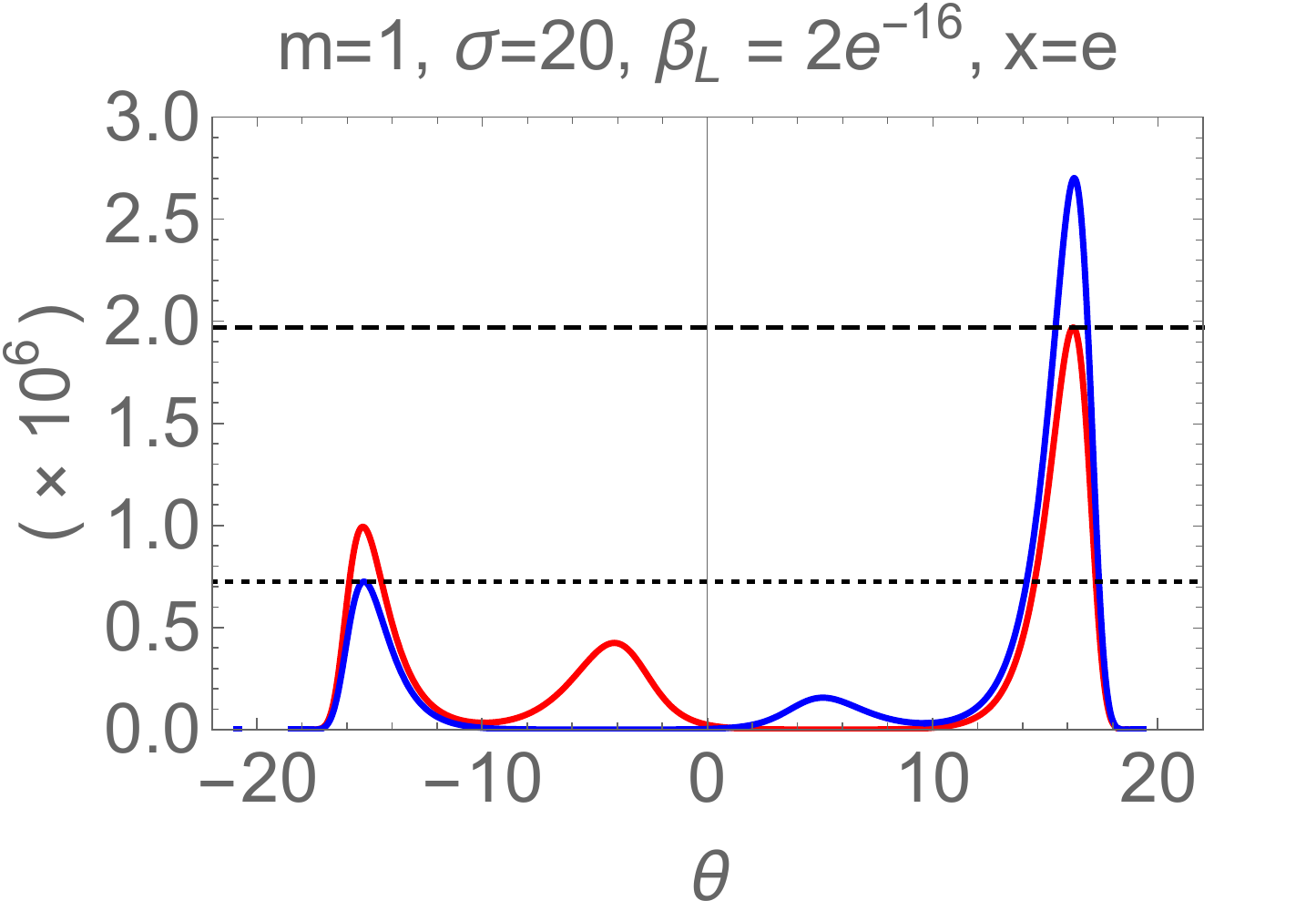} 
     \end{center} 
 \caption{Spectral Densities for $x=e$ ($T_L>T_R)$ and three values of the inverse temperatures ($\beta_L=2 e^{-3}, 2 e^{-11}$ and $2 e^{-16}$). For the two highest temperatures the vertical axis labels should be multiplied by $10^3$ and $10^6$, respectively, as indicated. In all panels, dashed (dotted) horizontal lines indicate the height of the free-fermion peaks, $0.04431.../\beta_{L}$ ($0.04431.../\beta_{R}$). A more complete picture of the dynamics can be gained from \href{https://youtu.be/suIftU1oNcw}{this video \cite{video4}}.}
 \label{ooeparden}
 \end{figure}

Let us now discuss how the spectral densities change in an out-of-equilibrium situation. Fig.~\ref{ooeparden} shows three examples for low, intermediate and large temperature which can be easily compared with Fig.~\ref{denprof}. 
We notice the following new features:
\bi
\item {\bf Free Fermion Regime:} For low temperatures (Fig.~\ref{ooeparden}, left panel) we recover the out-of-equilibrium free fermion behaviour
$$
\rho_p(\theta;\pm)=\frac{1}{2\pi}\cosh\theta \left(\frac{\Theta(\theta)}{1+e^{\beta_L\cosh\theta}}+ \frac{\Theta(-\theta)}{1+e^{\beta_R\cosh\theta}}\right)\,.
$$
The maxima are centered around $\theta=\log\frac{\beta_R}{2}$ and $\theta=-\log\frac{\beta_L}{2}$ and continue to be so even at higher temperatures. 

\item {\bf Intermediate Temperatures:} As in the equilibrium situation, the heights of the free fermion peaks start to change after the onset of the unstable particle (Fig.~\ref{ooeparden}, middle panel). However still the right peak of particle $+$ density coincides with the free fermion peak at temperature $\beta_L$ and the left peak of the $-$ spectral density coincides with the free fermion peak at inverse temperature $\beta_R$. These are the free fermion peaks that we had identified in the equilibrium situation. The opposite peaks, which have higher heights than they would in a free fermion theory, are the interacting peaks, as also identified in the equilibrium situation. Importantly, by contrast here the peaks of $+$ and $-$ particles have different heights.

\item{\bf Three Local Maxima:} For very high temperatures (Fig.~\ref{ooeparden}, right panel), we observe once more a structure with three local maxima per density. The additional (smaller) maxima are located at $-\log\frac{\beta_L}{2}-\sigma$ (red curve, $+$ spectral density) and $\log\frac{\beta_R}{2}+\sigma$ (blue curve, $-$ spectral density). Following the nomenclature used in the equilibrium situation, these are the subsidiary peaks. Once more, the excess area of the left-most, interacting peak in the density of $+$ particles (compared to the free fermion peak at inverse temperature $\beta_R$) roughly coincides with the area of the subsidiary peak in the density of $-$ particles. This is made more precise at the end of this subsection.

\begin{figure}[h!]
 \begin{center} 
 %\includegraphics[width=15.5cm]{eqcomparison.pdf} 
 %\hspace{-1.3cm}
 \includegraphics[width=16 cm]{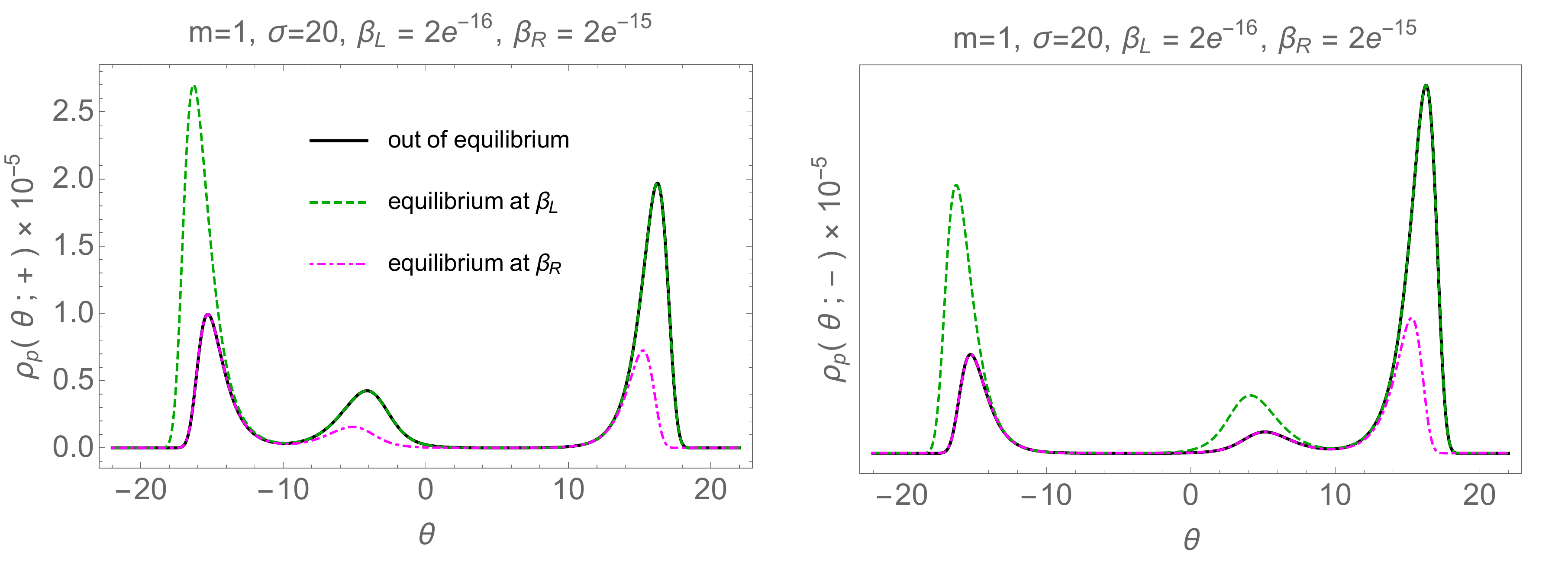}
     \end{center} 
 \caption{Spectral densities at equilibrium for temperatures $\beta=\beta_R=2 e^{-15}$ and $\beta=\beta_L=2 e^{-16}$ and out of equilibrium for the same temperatures.}
 \label{eparden}
 \end{figure}

\item {\bf Formation of the Unstable Particle:} As for the equilibrium case we can argue that the excess density of the interacting peak of the  $+$ spectral density ``couples" to the subsidiary peak of the $-$ spectral density and viceversa to form a finitely-lived unstable particle. The only difference with respect to the equilibrium case is that the areas and heights of  all  six maxima in  the two spectral densities are distinct.  In particular, the smallest maxima of both distributions are now different as one is governed by the right temperature and the other by the left temperature. This can be seen more precisely in the additional Fig.~\ref{eparden}. By computing the areas of all the peaks and comparing them to each other, this out-of-equilibrium analysis confirms the dynamical explanation of the formation of unstable particles, by allowing for an unambiguous identification of the coupling between $+$ and $-$ particles. A numerical evaluation of these areas is presented below.

\item {\bf Comparison to Equilibrium:}  Considering the densities in Fig.~\ref{eparden} we observe the following: for the $+$ particles density we find that the two right-most peaks -- the free fermion and subsidiary peaks -- are perfectly well described by the equilibrium density at inverse temperature $\beta=\beta_L=2 e^{-16}$ whereas the left-most peak -- the interacting peak -- is described by the equilibrium density at inverse temperature $\beta=\beta_R=2 e^{-15}$. The same ``cut and paste" structure is observed for the $-$ particles distribution, where the ``cut" is now located around $\theta=10$ (this is the the value of $\theta_0^{-}$ as seen from Fig.~\ref{theta*}). This behaviour can be best explained when matching densities with effective velocities. The velocities associated to the various types of peaks (free fermion, subsidiary and interacting) are distributed as in the equilibrium case, but now, these determine the initial bath the particles come from, and thus the temperature they carry. See the discussion in subsection \ref{suboutfull}.

\ei
Before concluding this subsection we would like to make our statements about the areas of the various maxima of the spectral densities a little bit more precise. For this purpose let us define the following quantities:
\begin{eqnarray}
   &&  A_+:= \int_{-R}^{t^+_{\rm{min}}} d\theta \, ( \, \rho_p ( \theta , + ) - \rho^{FF}_p ( \theta , +)_{\beta_L} \, ) \,, \qquad A_-:=\int_{t^-_{\rm{min}}}^R d\theta \, ( \, \rho_p ( \theta , - ) - \rho^{FF}_p ( \theta , - )_{\beta_R} \, ) \; , \nonumber\\
    &&  B_+ :=\int_{t^+_{\rm{min}}}^5 d\theta \,  \rho_p ( \theta , + ) \,, \qquad \qquad \qquad \qquad \qquad 
    B_ -:=\int_{-5}^{t^-_{\rm{min}}} d\theta \, \rho_p ( \theta , - )  \,.
\end{eqnarray}
where $R = \log 2/ \beta_L + 6$, $t^\pm_{\rm min}$ is the position of the local minimum of the spectral density that is located between the interacting and subsidiary peaks (that is approximately $\pm 10$ in Fig.~\ref{eparden}).  $\rho^{FF}_p(\theta;\pm)_\beta$ is the free fermion spectral density given by (\ref{ffparden}) at inverse temperature $\beta$. The subsidiary peaks of the $\pm$ spectral densities are then located approximately in the intervals $[t^+_{\rm{min}},5]$ and $[-5,t^-_{\rm{min}}]$. The choice of the integration limits is of course slightly arbitrary, so the areas below are just an illustration of the general statement that $A_+\approx B_-$ and $A_-\approx B_+$. In contrast to the equilibrium case it is now clear that $A_+\neq B_+$ and $A_-\neq B_-$, therefore our argument based on attributing a certain area of the spectral density curves to the formation of unstable particles is only plausible if we ``couple" the $\pm$ spectral density curves. 
\begin{table}[h!]
\centering
 \begin{tabular}{||c c c c c c c||} 
 \hline
  %\, & \, &  \, & \, & \, & \, & 
  $\log \beta_L/2$ & $t^+_{\rm min}$&  $t^-_{\rm min}$& $A_+$ & $B_+$ & $B_-$ & $A_-$ \\ [0.5ex] 
 \hline\hline
 $-15$ & $-9.9099$ & $9.4771$ & $26433.3$ & $68869.5$ & $24477.7$ & $70810.8$ \\
 $-16$ & $-9.9802$ & $9.6275$ & $70689.5$ & $189237.$ & $68793.3$ & $191195.$ \\
 $-17$ & $-9.9802$ & $9.7162$ & $191302.$ & $516162.$ & $189231.$ & $518292.$ \\
 \hline
 \end{tabular}
 \caption{Excess areas of the interaction peaks of the spectral densities $A_\pm$ compared to the areas of the subsidiary peaks $B_{\pm}$. As expected  $A_+\approx B_-$ and $A_-\approx B_+$. }
\end{table}

\subsection{Particle Currents}\label{suboutpartcurrent}
\begin{figure}[h!]
 \begin{center} 
    \hspace{-0.3cm}
 \includegraphics[width=7.8cm]{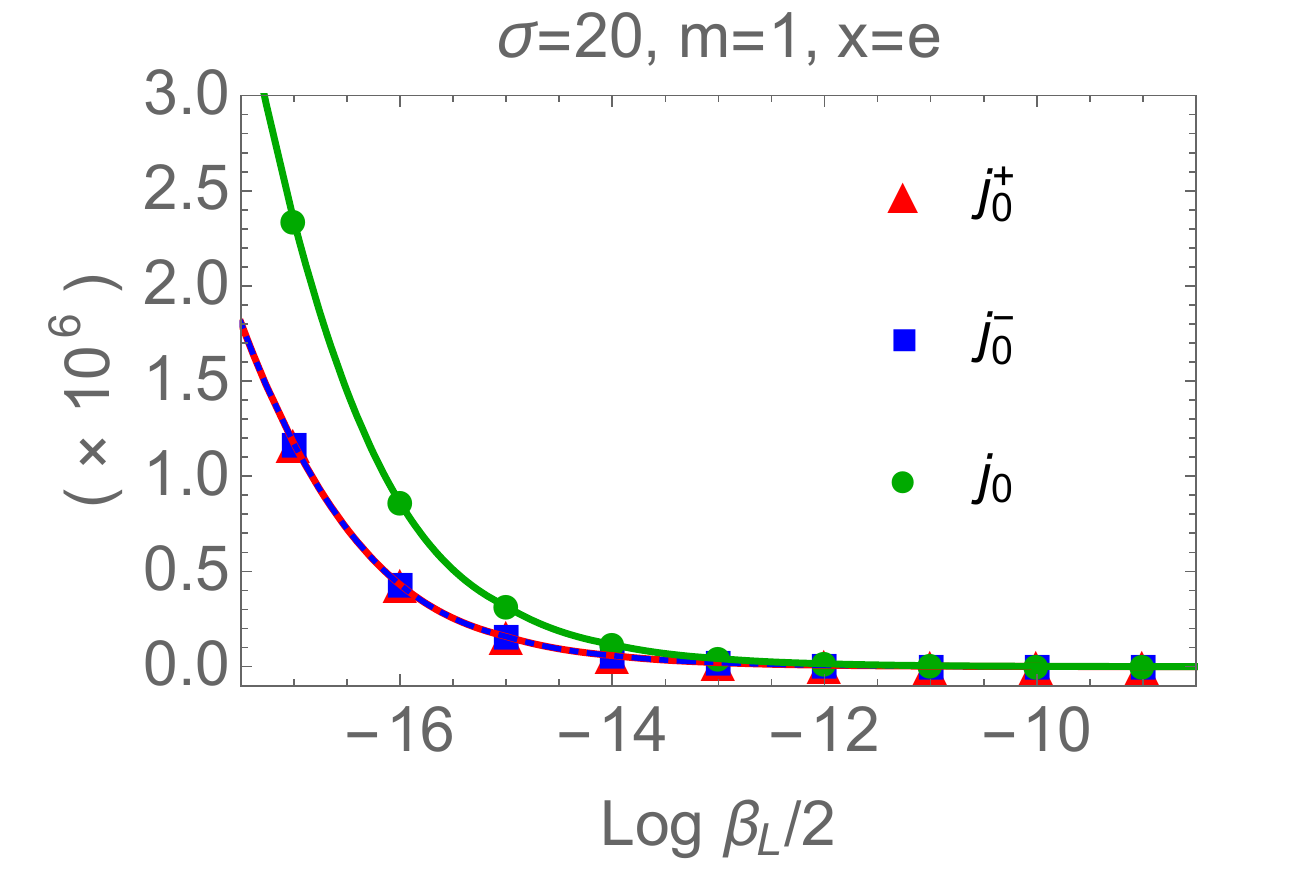} 
 \hspace{-0.3cm}
 \includegraphics[width=7.5cm]{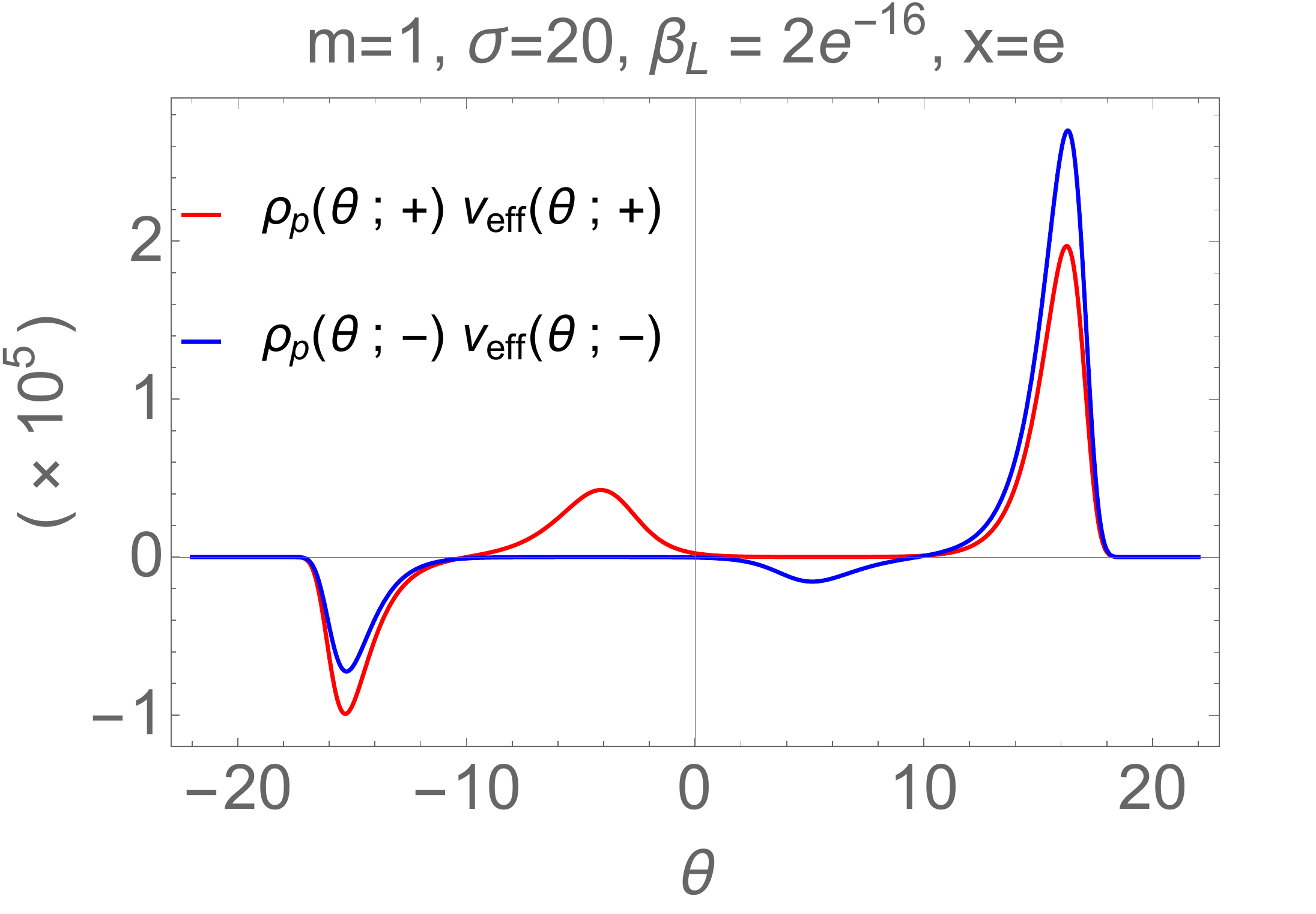} \\
 \includegraphics[width=7.8cm]{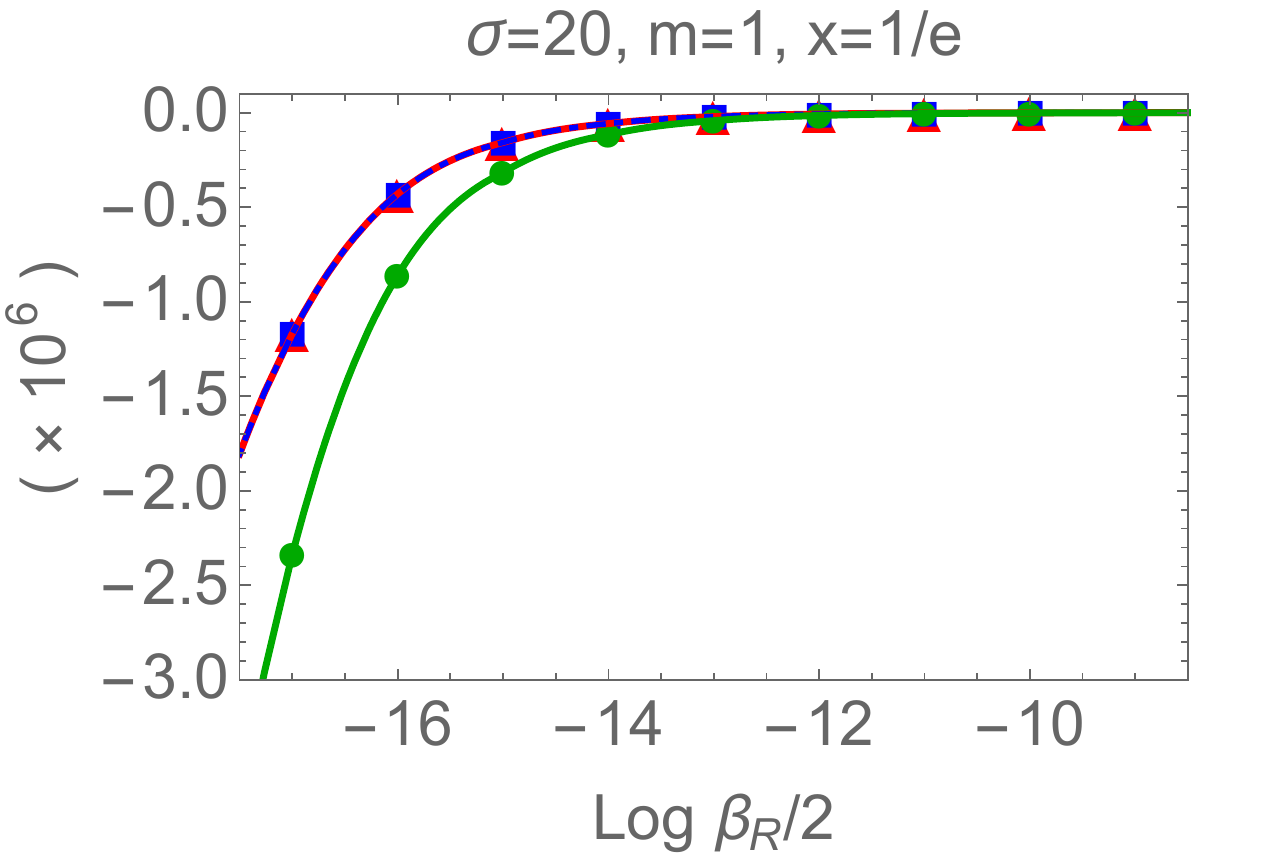}
 \hspace{-0.3cm}
  \includegraphics[width=7.8cm]{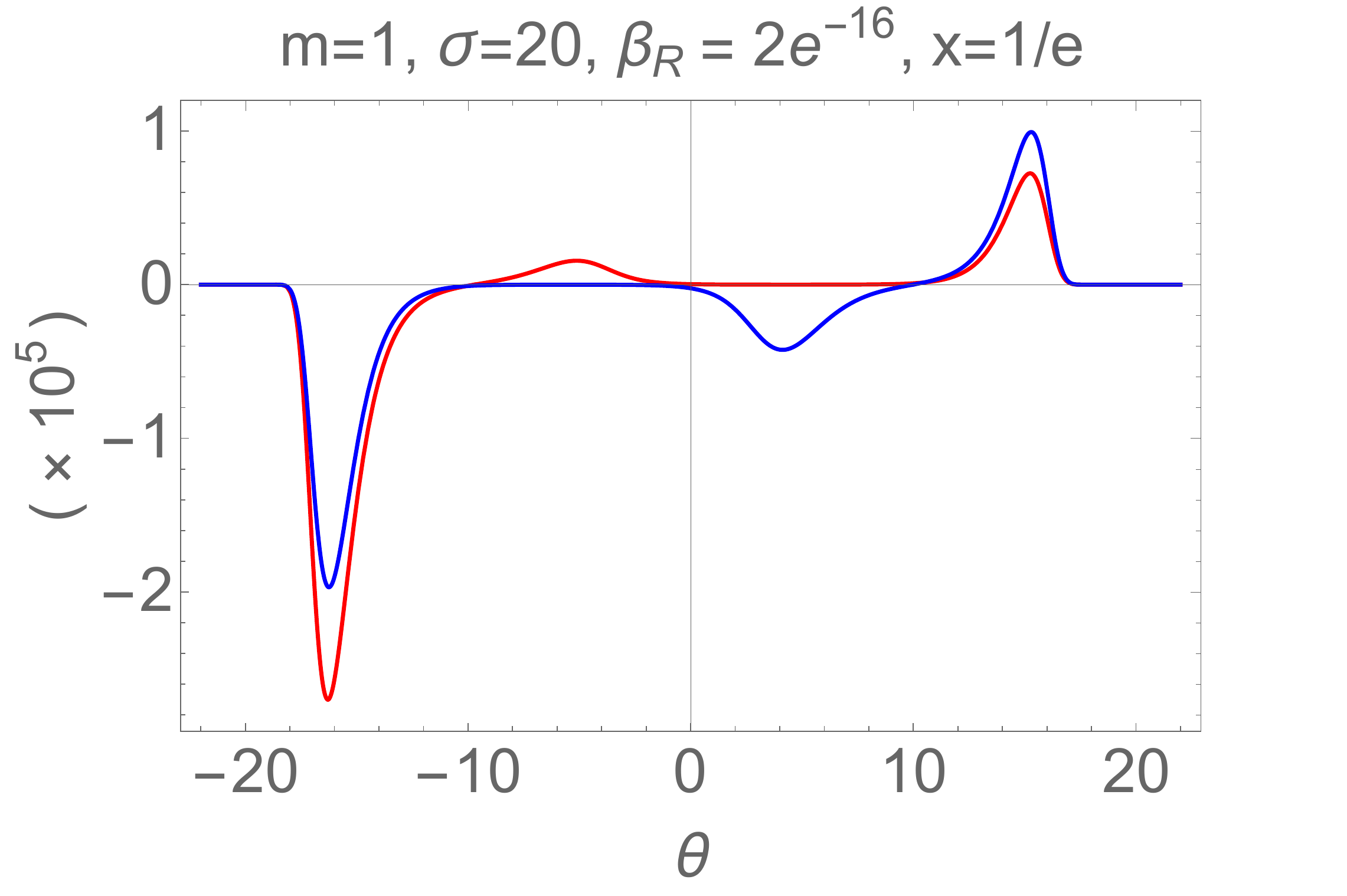} 
     \end{center} 
 \caption{Left panels: Total particle current $j_0$ (circles, green), particle current associated to particle $+$ (triangles, red) and to particle $-$ (squares, blue).  The vertical axis labels should be multiplied by a factor $10^6$, as indicated.  Right panels: The functions $\rho_p(\theta;\pm) v_{\rm{eff}}(\theta;\pm)$ whose integration gives the particle current. The vertical axis labels should be multiplied by factors $10^6$ and $10^5$, as indicated.}
 \label{pc}
 \end{figure}

The out-of-equilibrium particle currents differ substantially from their equilibrium values. They are presented in Fig.~\ref{pc}.
 The main features that we observe are the following:

\bi
\item {\bf (Almost) Identical Currents:} In contrast to the equilibrium picture, particle currents out of equilibrium are large, have no intricate structure and seem identical for both particle types. 
The reason for this is that the currents were very small at equilibrium and as soon as a temperature gradient is created, the current generated by this gradient is much larger than any previously existing current. Thus, contrary to the energy currents where the competition between equilibrium and out-of-equilibrium dynamics can be detected, for the particle currents, the out-of-equilibrium dynamics overwhelms any existing equilibrium current. In fact, we know from our numerical results that the currents are not exactly identical, their difference being of the order of the equilibrium particle currents. We can also see that the functions from whose integration they result (right panels) are not related by any obvious symmetry.

\item {\bf Sign of the Currents: } Currents are positive for $x>1$ and negative for $x<1$ as governed by the temperature gradient. In this case the signs of particle and energy currents coincide. A simple justification for this behaviour can be found again from the spectral densities and effective velocities. The sign can also be worked out from the functions on the right panels. Contrary to the equilibrium case, they are no longer almost odd functions. Instead the areas above and below the horizontal axis are visibly different. 
\ei
\subsection{Out-of-Equilibrium Full Dynamics}\label{suboutfull}

As for the equilibrium case let us once more contrast the behaviour of velocities with that of densities. The picture is similar to equilibrium (see Fig.~\ref{extra}).
 In particular, we can make the same arguments about the formation of the unstable particle.  There are however some new properties worth mentioning: 
 
 \bi 
\item {\bf Following the Original Baths:} Let us compare Fig.~\ref{ooeparden}, right panel and the corresponding effective velocities, which would look very much like Fig.~\ref{ooeveff}, top right panel. Consider the $+$ spectral density.  As for the equilibrium situation we find that the two right most peaks in the density (centered around rapidities $16,- 4$) are associated with velocity $+1$ whereas the left-most peak (at rapidity $-15$) has velocity $-1$. Thus particles distributed around rapidities $16, -4$ are moving from left to right. This means that they were originally thermalized on the left bath with inverse temperature $\beta_L$. This is the reason why their density coincides with the equilibrium density at inverse temperature $\beta_L$ as shown in Fig.~\ref{eparden}. Similarly, particles with rapidity around -15 are moving from right to left and therefore were thermalized on the right bath at inverse temperature $\beta_R$.  The same kind of argument can be made for the $-$ particle. 
\item {\bf Currents:} We have previously observed that, contrary to the equilibrium case, both the particle and energy currents are now positive for $x>1$ and negative for $x<1$.  Indeed, we now find that the sum of the areas of the two right-most peaks of the $+$ spectral density is larger than the area of the left-most peak. So there is a much larger density of particles with velocity $+1$ than there is with velocity $-1$ and a large positive particle current is generated as a result. Similarly, there is a larger density of $+$ particles around rapidity $16$ than there are around rapidity $-15$ so there are more highly energetic particles with positive than negative momentum and these produce a net positive energy current. As usual, a similar argument can be made for the $-$ particles.

\ei

\section{Application: Connecting CFTs with Different Central Charges}

Many times in this paper we have highlighted the property that for temperatures well below the unstable particle mass our model describes a pair of free Majorana fermions whereas for temperatures  well above the unstable particle mass a new critical point is reached with central charge $1.2$. This means that our model provides an ideal opportunity to investigate the properties of the energy current in the conformal regime when two theories of different central charges are connected. 

In order to carry out this experiment we need $\beta_L$ and $\beta_R$ to differ by many orders of magnitude for the two halves to be deep into the two conformal regimes. Looking for instance at Fig.~\ref{difT} for the currents and densities or at Fig.\ref{cfun} for the $c$-function, we see that for $\sigma=20$  we generally need one inverse temperature to be much smaller than $\ll2 e^{-10}$ and the other much larger than the same value, ideally towards the middle of each plateau. 

Table~\ref{ourtable} shows various pairs of possible choices together with the estimated values of the current, the density and a coefficient $a$ which is defined by
\beq
|j_1|=\frac{a\pi}{12} T^2,
\eeq
where $T$ is the largest temperature.  As we can see in the table,  the value of $a$ is in all cases very close to the highest central charge $c=1.2$. As expected, we also see that the sign of the current is reversed when the choice of temperatures is exchanged (compare the first and fifth rows in the table). 
\begin{table}[h!]
\begin{center}
 \begin{tabular}{||c c c c c c||} 
 \hline
 $x$ & $\beta_L$ & $\beta_R$ & $a$ & $\beta^2 \,\texttt{j}_1$ & $\beta^2 \, \texttt{q}_1$ \\
 \hline\hline
% \, & \, &  \, & \, & \, & \\
$1.\times e^{-7}$ & $2.\times e^{-7}$ &  $2.\times e^{-14}$ & $1.1986$ & $-0.3138$ & $0.3139$ \\
$1.\times e^{4}$ & $2.\times e^{-14}$ &  $2.\times e^{-10}$  & $1.1986$ & $0.3137$ & $0.3140$ \\
$1.\times e^{5} $& $2.\times e^{-14}$ &  $2.\times e^{-9}  $& $1.1986$ & $0.3138 $& $0.3139$ \\
$1.\times e^{6} $& $2.\times e^{-15}$ &  $2.\times e^{-9}  $& $1.1988$ & $0.3139 $& $0.3139$ \\
$1.\times e^{7} $& $2.\times e^{-14}$ &  $2.\times e^{-7} $& $1.1986$ & $0.3138 $& $0.3139 $\\
$1.\times e^{8} $& $2.\times e^{-15}$ &  $2.\times e^{-7} $& $1.1988$ &$ 0.3139 $& $0.3139$ \\
$1.\times e^{9} $& $2.\times e^{-16}$ &  $2.\times e^{-7} $& $1.1991$ & $0.3139 $& $0.3139$ \\
 \hline
 \end{tabular}
 \end{center}
 \caption{Investigation of the energy current and energy density when connecting two CFTs of central charges $1$ and $1.2$. In the last two columns we chose $\beta=\min(\beta_L,\beta_R)$.}
 \label{ourtable}
\end{table}
As expected, the dependence on the lower temperature is negligible compared to the numerical error. From the numerical results we can postulate that the leading behaviour of the current and density is given by 
\beq
\texttt{j}_1\approx \frac{a \pi}{12}T_L^2\, \qquad \mathrm{for } \qquad T_L \gg T_R \qquad \mathrm{and} \qquad \texttt{j}_1=-\frac{a \pi}{12}T_R^2\qquad \mathrm{for} \qquad T_R\gg T_L\,,
\eeq
and 
\beq
\texttt{q}_1\approx \frac{a \pi}{12}T_L^2\, \qquad \mathrm{for } \qquad T_L \gg T_R \qquad \mathrm{and} \qquad \texttt{q}_1=\frac{a \pi}{12}T_R^2\qquad \mathrm{for} \qquad T_R\gg T_L\,,
\eeq
with $a=\frac{6}{5}$.

The problem of connecting two critical theories with different central charges has been studied in several previous works. However, none of these works has considered a situation that is directly comparable to ours and indeed they all predict a different behaviour of the current.  
For instance in \cite{BDV,FKSF} the connection of two free CFTs of different central charges was considered, whereas in 
 \cite{MVCRD} two different critical models are connected by a defect which breaks conformal invariance.   However, the argument put forward in \cite{FKSF} about a possible ``bottleneck" effect whereby the smallest central charge limits the growth of the currents, seems rather plausible and yet does not hold here.  A potential explanation is that with an actual, localised impurity in the dynamics, a true bottleneck effect may arise where the impurity is unable to carry more degrees of freedom through it than those supported by the theory with the smallest central charge. By contrast, in the present setup, the middle region where nonequilibrium currents build up may be very extended, and can accumulate large amounts of energy. We do not have a full understanding at this stage, but it will be interesting to explore this problem further. 

\section{Conclusion}

In this paper we have applied the GHD framework to the study of an integrable relativistic quantum field theory known as the $SU(3)_2$-Homogeneous sine-Gordon model.
The model is of interest in the GHD context because it has two novel features: it  has two stable and one unstable particle and its (diagonal) scattering matrix breaks parity, meaning that $S_{+-}(\theta)\neq S_{-+}(-\theta)$ where $\pm$ are the two stable particle types. The effect of both features on all dynamical properties of the model, including currents and densities, turns out to be quite dramatic. 

\medskip

Generically, the presence of the unstable particle means that there are three interesting temperature regimes: (1) At low temperatures, the unstable particle cannot be formed and the two stable modes effectively behave as free Majorana fermions, giving the known results and restoring parity; (2) At sufficiently large temperatures, the unstable particle is present giving rise to a new plateau in some quantities, such as the temperature-scaled energy currents, or to a change to the structure of the local maxima of the spectral densities; (3) At very high temperatures a new fixed point is reached and the conformal behaviour is recovered for central charge $c=\frac{6}{5}$.  These three regions are found both in the equilibrium and out-of-equilibrium dynamics.

\medskip

One of the most surprising results is that even at equilibrium it is possible to speak of a rich and interesting dynamics. Such dynamics can be explored specially by looking at the individual contributions of the stable particles to currents and densities and by studying the properties of their effective velocities and density distributions. Perhaps the most interesting finding is the identification of the dynamical process by which the unstable particle is formed: for high enough temperatures (i.e. commensurate with the unstable particle's mass) the density of $\pm$ particles is increased for some range of rapidities; this ``excess" spectral density is such that increases for $\pm$ particles can be matched and interpreted as the spectral density of unstable particles. In addition, the excess spectral densities of $\pm$ particles arise for rapidities for which their respective effective velocities are identical. As a result, the unstable particle can be seen as finitely-lived bound state of co-propagating particles $(+-)$ or $(-+)$; these bound states are continuously replenished thanks to the high temperature and the availability of co-moving particles of opposite types, and thus are found in stable proportion in the thermal bath. 

\medskip

Out of equilibrium, we observe that many equilibrium features are preserved while others are modified: energy and particle currents are positive for $T_L>T_R$ and negative otherwise, indicating that the temperature gradient (rather than the equilibrium dynamics) is the leading force in determining the out-of-equilibrium dynamics. A consequence of parity breaking and the unusual features of the phase shifts is that particle $+$ is particularly sensitive to changes in $\beta_L$ whereas particle $-$ is most sensitive to changes in $\beta_R$. Analysing the spectral densities in conjunction with the effective velocities at very high temperatures we observe that values of $\theta$ for which the velocities are $+1 (-1)$ can be matched to spectral densities that reproduce the equilibrium profile on the left (right) reservoir, where the particles were originally thermalized. Unstable particles are still formed by the combinations $(+-)$ or $(-+)$ of pairs propagating at the same speed, but in the out-of-equilibrium set-up the density of each pair-type is distinct. 

\medskip

A by-product of our analysis is the opportunity to address the question: what is the out-of-equilibrium dynamics following the connection of two thermal CFTs of different central changes?
The special nature of the current model allows us to engineer such a partitioning protocol by setting the right and left baths at very low and very high temperatures, respectively. By doing so we will have on one bath two free Majorana fermions with total central charge $c=1$ and on the other bath an interacting CFT with central charge $c=\frac{6}{5}$. Our  numerical results suggest 
\beq
|\texttt{j}_1|=q_1\approx \frac{a \pi}{12}T^2\, \qquad \mathrm{with} \qquad a=\frac{6}{5}\,,
\eeq
and $T$ is the largest temperature. This is different from any of the formulae found in \cite{BDV,MVCRD,FKSF} and likely due to our special set-up. A better understanding of the conditions under which each behaviour is to be expected is still needed. 
 
 \medskip
 
It would be interesting to study other theories in this family, where more than one unstable excitation is present, but we expect that the main physical picture will remain the same. We would also like to explore other quench protocols where the effect of the unstable particle may be different. Most interesting perhaps would be to have a lattice or cold-atom realization of these or similar models, paving the way towards a study of unstable particles, parity breaking, out-of-equilibrium dynamics and integrability in an experimental set up. 

 \medskip
  \medskip

\noindent {\bf Acknowledgements:} The authors thank Jacopo Viti for discussions on the work \cite{MVCRD}.  BD acknowledges funding from the Royal Society under a Leverhulme Trust Senior Research Fellowship, ``Emergent hydrodynamics in integrable systems: non-equilibrium theory", ref.\\ SRF\textbackslash R1\textbackslash 180103.

\appendix 

\section{Temperature Dependence of $\theta^\pm_0$}
In this section we present some results for the values of $\theta^\pm_0$, that is the discontinuity of the occupation numbers $n(\theta;\pm)$ in the non-equilibrium steady state.
\begin{figure}[h!]
 \begin{center} 
 \includegraphics[width=8cm]{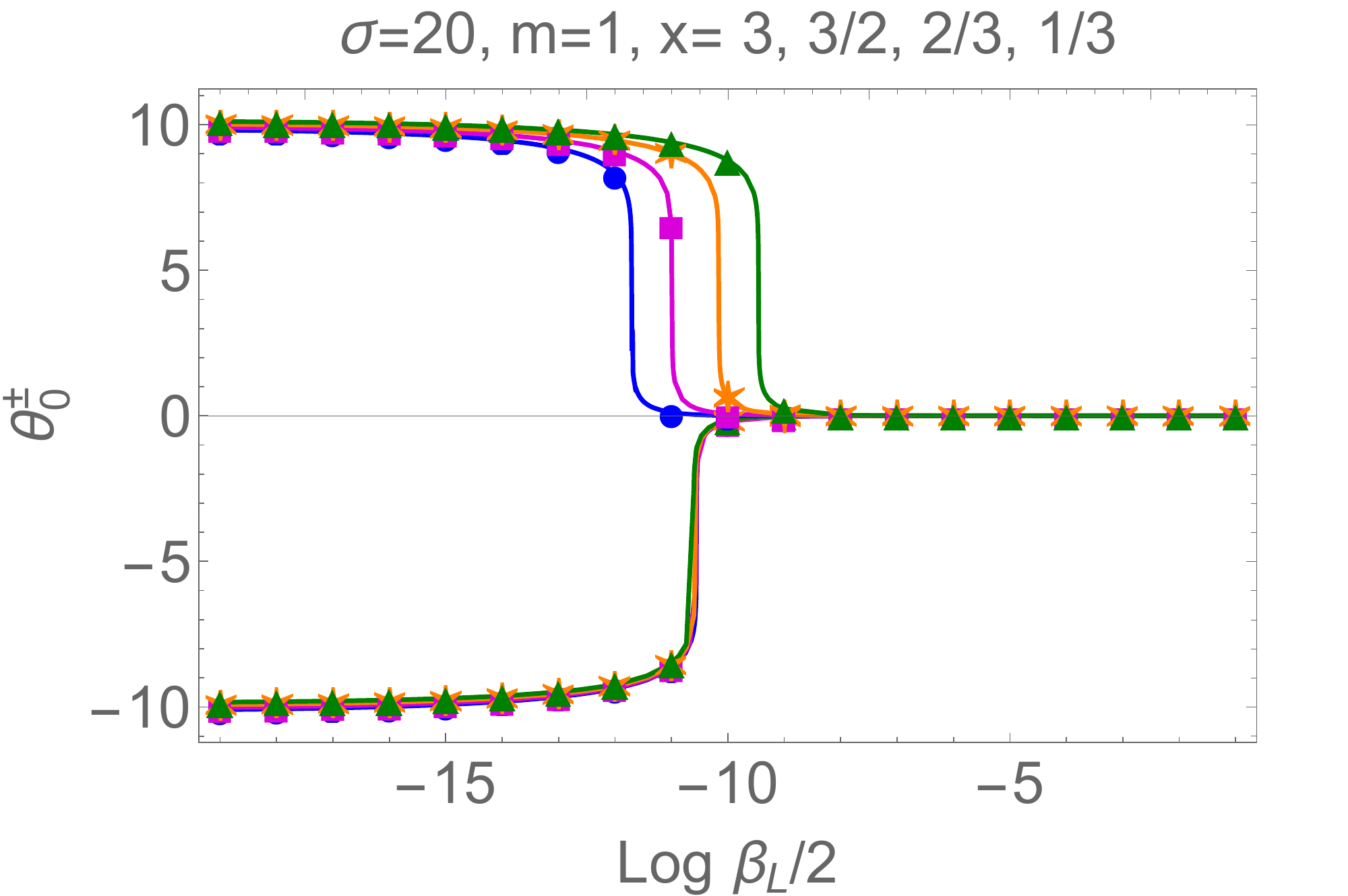} 
    \end{center} 
 \caption{Values of $\theta_0^\pm$ for various ratios $x$ as functions of $\log\frac{\beta_L}{2}$, with $\theta_0^+\leq 0$ and $\theta_0^-\geq 0$. The signs are as expected from the behaviours of the effective velocities seen in Fig.~\ref{ooeveff}.
 The same symbols are employed for  $\theta_0^\pm$ and each given $x$. } 
 \label{theta*}
\end{figure}
We observe the following main features:
\bi
\item  {\bf Temperature Dependence:} As already observed for other quantities, we see how the value $\theta_0^+$ is almost entirely determined by the value of $\beta_L$ whereas $\theta_0^-$ is very sensitive to changes in $\beta_R$.

\item {\bf Free Fermion Regime:} Since $\sigma=20$ for $\log\frac{\beta_{L,R}}{2}>-10$ the temperatures are too low for the unstable particle to be present and we find $\theta_0^\pm=0$, that is the free fermion result.

\item {\bf Unstable Particle Onset:} A marked change in behaviour is observed at $\log\frac{\beta_{L,R}}{2}=-10$ with the onset of a plateau. For $\log\frac{\beta_{L,R}}{2}<-10$ we find $\theta_0^+=-\theta_0^-\approx -10$. This value $-10$ is again related to the value of $\sigma$. Although the resulting functions seem almost discontinuous (i.e~ step-functions) we have no reason to think that this is the case. 

\item {\bf Zeroes of the Effective Velocities:} We observed in subsection \ref{sooeveff} that for some values of $\beta_L, \beta_R$ the effective velocities develop an intermediate plateau of height zero. Thus they appear to have a continuous set of zeroes, corresponding to a continuous set of values of $\theta_0^\pm$. This is a very unusual phenomenon, as far as we know, not seen previously. The shape of the functions $\theta_0^\pm$ shows us that this occurs precisely when $\log\frac{\beta_{L}}{2} \approx-10$ for particle $+$ and when $\log\frac{\beta_{R}}{2} \approx-10$ for particle $-$ that is, at the onset of the unstable particle. As mentioned in the previous point, our understanding is that the values of $\theta_0^\pm$ are always unique but that for some small range of temperatures our algorithm is not accurate enough to precisely identify these values. In other words, the intermediate plateau of the effective velocities is never exactly flat, but for temperatures $\log\frac{\beta_{L,R}}{2} \approx-10$ its slope is too small to be seen numerically.
 \ei

\section{Numerical Recipe}

In this appendix we discuss briefly some details of the Mathematica programme we have used to generate all the numerical results presented in this paper. As usual in the TBA context, the TBA/GHD equations are solved numerically starting with a discretization of the variable $\theta$ within a finite interval. For this we exploit a well-known property of all relevant TBA functions namely, that they double-exponentially fall off for rapidities larger than $\log 2/\beta$ or smaller than $\log \beta/2$ (and similarly in the out-of-equilibrium situation). In our numerics we have chosen a slightly larger interval $\left[ \log \beta /2  - \sigma/4 \, , \,  \log 2/ \beta + \sigma/4 \right]$ which grows with temperature. In the out-of-equilibrium regime we choose $\beta$ to be the inverse of the highest temperature. 

We have kept the number $M$ of discrete equidistant rapidity values fixed. It is clear that the larger $M$ is,  the better the approximation to the continuum. However, a very large $M$  increases drastically the running time of the programme. In all our numerical analysis we have set $M =200$. This value has been chosen in such a way as to ensure that a number of benchmark results are reproduced. For instance, we reproduce the  expected pattern of the $c$-function at equilibrium (see Fig.~\ref{cfun}) as well as the known free Majorana fermion results in the relevant temperature range, both at and out of equilibrium. 

We have focussed on studying the temperature-dependence of the TBA quantities described in Section 2 exactly in the middle of the light-cone (so, for ray $x/t=0$). For simplicity, we have set the parameters of the theory as  $m=1$ and $\sigma = 20$.  We can summarise the algorithm we have implemented as follows:
\begin{itemize}
\item[(a)] For fixed values of $\beta_{L,R}$ solve (\ref{pseudoen}) for the left and right steady states and compute  $n^R(\theta;\pm)$ and $n^L(\theta; \pm)$ using (\ref{ndefini}). 
\item[(b)] Solve (\ref{dress}) for $h_i(\theta;\pm)=p
(\theta;\pm)$ recursively. Start by setting $\theta_0^\pm$ in (\ref{on}) to some trial value (say 0). Solve recursively for $p^{\mathrm{dr}}(\theta;\pm)$ until convergence is achieved. 
\item[(c)] Once a solution for  $p^{\mathrm{dr}}(\theta;\pm)$  has been obtained, find the solution to $p^{\mathrm{dr}}(\theta;\pm)=0$. This will give a new value of $\theta_0^\pm$. 
\item[(d)] Repeat (b) and (c) with this new value of $\theta_0^\pm$ and again as many times as necessary until a stable value of $\theta_0^\pm$ is reached.
\item[(e)] Employ the solution (\ref{on}) to evaluate any dressed quantity of interest $h^{\mathrm{dr}}(\theta;\pm)$.
\item[(f)] Evaluate (\ref{q}) and (\ref{j}).
\item[(g)] Repeat for a different right- and left-temperatures. %and/or $\sigma$. In this paper we have focussed on the ray $\zeta=0$ and the value $\sigma=20$.
\end{itemize}
In (c) and (e) the convergence of the dressing operation is ensured by the condition that the difference of the outcome given by the last iteration and the preceding one is smaller than the module of a certain number $\alpha$. In all of the cases, $\alpha$ has been chosen to be no larger than $10^{-4}$ generally ensuring very high precision. Similar arguments hold for the convergence of (a) and (d). \\

%We have checked the consistency of our numerical results in several ways. For instance, strong evidence is provided by the very good agreement with the CFT values of energy currents and energy densities at either very high and very low temperatures as shown in subsection 3.1 and subsection 4.1.

%\bibliography{Ref}
%\bibliographystyle{phreport}

\end{document}